\documentclass[longauth]{aa}  

\usepackage{graphicx}
\usepackage{txfonts}

\usepackage{xspace}
\usepackage[sticky-per=false,
    separate-uncertainty,
    print-unity-mantissa=false,
    separate-uncertainty-units=single,
    angle-symbol-over-decimal,
    ]{siunitx}
\newcommand{\lya}{Ly$\alpha$\xspace}
\newcommand{\halpha}{H$\alpha$\xspace}
\newcommand{\hi}{H\,{\sc i}\xspace}
\newcommand{\hii}{H\,{\sc ii}\xspace}
\newcommand{\ci}{[C\,{\sc i}]\xspace}
\newcommand{\ciim}{[C\,{\sc ii}]\,\SI{158}{\micro\meter}\xspace}
\newcommand{\cii}{[C\,{\sc ii}]\xspace}
\newcommand{\cplus}{C\textsuperscript{+}\xspace}

\newcommand{\heii}{He\,{\sc ii}\xspace}
\newcommand{\civ}{C\,{\sc iv}\xspace}

\newcommand{\sysname}{J1000+0234\xspace}

\newcommand{\ca}{CRISTAL-01a\xspace}
\newcommand{\cb}{the DSFG\xspace}

\renewcommand{\arraystretch}{1.1}
\DeclareSIUnit\jansky{Jy}
\DeclareSIUnit\erg{erg}
\DeclareSIUnit\mag{mag}
\DeclareSIUnit\parsec{pc}
\DeclareSIUnit\arcsec{arcsec}
\DeclareSIUnit\arcsect{^{\prime\prime}}
\DeclareSIUnit\msun{M_\odot}
\DeclareSIUnit\lumsol{L_\odot}
\DeclareSIUnit\zsun{Z_\odot}
\DeclareSIUnit\year{yr}
\DeclareSIUnit\beam{beam}
\DeclareSIUnit\angstrom{\text {Å}}

\defcitealias{Capak2008ExtremeStarburst}{C08}
\defcitealias{Jimenez-Andrade2023LyaNebulaeGalaxyPair}{J23}
\defcitealias{GomezGuijarro2018AlmaMinorMergersSmgs}{GG18}

\usepackage{xcolor}
\usepackage{multirow}
\usepackage{natbib}
\bibpunct{(}{)}{;}{a}{}{,} 

\usepackage[colorlinks=true,
linkcolor=blue, citecolor=blue, filecolor=blue, urlcolor=blue,
unicode=true]{hyperref}

\begin{document}

\title{The ALMA-CRISTAL survey} 

   \subtitle{Discovery of a 15\,kpc-long  gas plume in a $z=4.54$ Lyman-$\alpha$ blob}

   \author{M. Solimano
  \inst{1}
  \and
  J. Gonz\'alez-L\'opez\inst{2,1}
  \and
  M. Aravena\inst{1}
  \and
  R. Herrera-Camus\inst{3} 
  \and
  I. De Looze\inst{4, 5}
  \and
  N. M. F\"orster Schreiber\inst{6}
  \and
  J. Spilker\inst{7}
  \and
  K. Tadaki\inst{8}
  \and
  R. J. Assef\inst{1}
  \and
  L. Barcos-Mu\~noz\inst{9,10}
  \and
  R. L. Davies\inst{11, 12}
  \and
  T. D\'iaz-Santos\inst{1, 13, 14}
  \and
  A. Ferrara\inst{15}
  \and
  D. B. Fisher\inst{11, 12}
  \and
  L. Guaita\inst{16}
  \and
  R. Ikeda\inst{17, 18}
  \and
  E. J. Johnston\inst{1}
  \and
  D. Lutz\inst{9}
  \and
  I. Mitsuhashi\inst{26, 18}
  \and
  C. Moya-Sierralta\inst{19}
  \and
  M. Rela\~no\inst{20, 21}
  \and
  T. Naab\inst{22}
  \and
  A. C. Posses\inst{1}
  \and
  K. Telikova\inst{1}
  \and
  H. {{\"U}bler}\inst{23,24}
  \and
  S. van der Giessen\inst{4, 25, 22}
  \and
  V. Villanueva\inst{3}
}

\institute{
  Instituto de Estudios Astrof\'isicos, Facultad de Ingenier\'ia y Ciencias, %
  Universidad Diego Portales, Av.  Ej\'ercito Libertador 441, Santiago, Chile [C\'odigo Postal 8370191] \email{manuel.solimano@mail.udp.cl} 
  \and
  Las Campanas Observatory, Carnegie Institution of Washington, %
  Casilla 601, La Serena, Chile
  \and
  Departamento de Astronom\'ia, Universidad de Concepci\'on,%
  Barrio Universitario, Concepci\'on, Chile
  \and
  Sterrenkundig Observatorium, Ghent University, Krijgslaan%
  281-S9, B-9000 Ghent, Belgium
  \and
  Department of Physics \& Astronomy, University College London, %
  Gower Street, London WC1E\,6BT, UK
  \and
  Max-Planck-Institut f\"ur Extraterrestiche Physik (MPE), %
  Giessen-bachstr., 85748, Garching, Germany
  \and
  Department of Physics and Astronomy and George P. and %
  Cynthia Woods Mitchell Institute for Fundamental Physics %
  and Astronomy, Texas A\&M University, College Station, TX, USA
  \and
  Faculty of Engineering, Hokkai-Gakuen University, Toyohira-ku, %
  Sapporo 062-8605, Japan
  \and
  Joint ALMA Observatory, Alonso de C\'{o}rdova 3107, Vitacura, %
  Santiago, Chile
  \and
  National Radio Astronomy Observatory, 520 Edgemont Road, %
  Charlottesville, VA 22903, USA
  \and
  Centre for Astrophysics and Supercomputing, Swinburne Univ. %
  of Technology, PO Box 218, Hawthorn, VIC, 3122, Australia
  \and
  ARC Centre of Excellence for All Sky Astrophysics in %
  3 Dimensions (ASTRO 3D), Australia
  \and
  Institute of Astrophysics, Foundation for Research and %
  Technology-Hellas (FORTH), Heraklion, 70013, Greece
  \and
  Chinese Academy of Sciences South America Center for %
  Astronomy (CASSACA), National Astronomical Observatories, %
  CAS, Beijing, 100101, PR China
  \and
  Scuola Normale Superiore, Piazza dei Cavalieri 7, %
  I-50126 Pisa, Italy
  \and 
  Universidad Andr\'es Bello, Facultad de Ciencias Exactas, %
  Departamento de F\'isica, Instituto de Astrof\'isica, %
  Fernandez Concha 700, Las Condes, %
  Santiago RM, Chile
  \and
  Department of Astronomical Science, SOKENDAI (The Graduate University for Advanced Studies), Mitaka, Tokyo 181-8588, Japan
  \and
   National Astronomical Observatory of Japan, 2-21-1 Osawa, %
   Mitaka, Tokyo 181-8588, Japan
  \and
  Instituto de Astrof\'isica, Facultad de F\'isica, %
  Pontiﬁcia Universidad Cat\'olica de Chile, Santiago 7820436, Chile
  \and
  Dept. Fisica Teorica y del Cosmos, Universidad de Granada, %
  Granada, Spain
  \and
  Instituto Universitario Carlos I de F\'{i}sica Te\'{o}rica %
  y Computacional, Universidad de Granada, %
  E-18071 Granada, Spain
  \and
  Max-Planck Institute for Astrophysics, %
  Karl Schwarzschildstrasse 1, 85748, Garching, Germany
  \and
  Cavendish Laboratory, University of Cambridge, 19 J.J. %
  Thomson Avenue, Cambridge, CB3 0HE, UK
  \and
  Kavli Institute for Cosmology, University of Cambridge, %
  Madingley Road, Cambridge, CB3 0HA, UK
  \and
  Leiden Observatory, Leiden University, NL-2300 RA Leiden, %
  Netherlands
  \and
  Department of Astronomy, The University of Tokyo, 7-3-1 %
  Hongo, Bunkyo, Tokyo 113-0033, Japan
}

   \date{Received -; accepted -}

   \abstract{
	   Massive star-forming galaxies in the high-redshift universe host large reservoirs of cold gas in their circumgalactic medium (CGM). Traditionally, these reservoirs have been linked to diffuse \hi Lyman-$\alpha$ (\lya) emission extending beyond $\approx$\SI{10}{\kilo\parsec} scales. In recent years, millimeter/submillimeter observations are starting to identify even colder gas in the CGM through molecular and/or atomic tracers such as the \ciim transition.
  In this context, we study the well-known \sysname system at $z=4.54$ that hosts a massive dusty star-forming galaxy (DSFG), a UV-bright companion, and a \lya blob.
   We combine new ALMA \cii line observations taken by the CRISTAL survey with data from previous programs targeting the \sysname system, and achieve a deep view into a DSFG and its rich environment at a \ang{;;0.2} resolution. 
   We identify an elongated \cii-emitting structure with a projected size of \SI{15}{\kilo\parsec} stemming from the bright DSFG at the center of the field, with no clear counterpart at any other wavelength. The plume is oriented $\approx$\ang{40} away from the minor axis of the DSFG, and shows significant spatial variation of its spectral parameters. In particular, the \cii  emission shifts from \SI{180}{\kilo\meter\per\second} to \SI{400}{\kilo\meter\per\second} between the bottom and top of the plume, relative to the DSFG's systemic velocity. At the same time, the line width starts at $400-\SI{600}{\kilo\meter\per\second}$ but narrows down to \SI{190}{\kilo\meter\per\second} at top end of the plume.  We discuss four possible scenarios to interpret the \cii plume: a conical outflow, a cold accretion stream, ram pressure stripping, and gravitational interactions. While we cannot strongly rule out any of these with the available data, we disfavor the ram pressure stripping scenario due to the requirement of special hydrodynamic conditions.
   }


   \titlerunning{Discovery of a gas plume in a \lya blob}
   \authorrunning{Solimano et al.}

   \maketitle
%

\section{Introduction}\label{sec:intro}

The multiphase gas envelope around galaxies, known as the circumgalactic medium \citep[CGM; e.g.,][]{Tumlinson2017CircumgalacticMediumReview}, plays a major role in galaxy growth and evolution. In particular, its cool phase ($T\lesssim \SI{1e4}{\kelvin}$) hosts the gas reservoirs needed to fuel star formation and supermassive black hole accretion. The regulation of these activities requires cool gas to be transported in and out of the interstellar medium (ISM), in what is called the baryon cycle \citep[e.g.,][]{PerouxHowk2020CosmicBaryonAndMetalCycles}, a process where the CGM is the main stage. For example, the CGM is the site where galactic scale outflows expand into \citep[e.g.,][]{Veilleux2020CoolOutflowsReview}, and where narrow, cold accretion streams flow into galaxies \citep[e.g.,][]{Dekel2009ColdModeAccretion}.

Observationally, cool CGM gas manifests in several ways. For example, it is routinely detected as \hi and/or metal absorption against the backlight of distant quasars at any redshift \citep[e.g.,][]{Werk2013CosHalos, ZhuAndMenard2013SdssMgiiAbsorbers, Turner2014MetalLineAbsorptionKeck}.
In recent years, high metallicity ($[\mathrm{M}/\mathrm{H}]\approx-1$), high column density ($N_\mathrm{H} \gtrsim \SI{2e20}{\per\centi\meter\squared}$) absorbers at $z\gtrsim4$ have been found at distances as large as \SIrange{15}{45}{\kilo\parsec} from their host galaxies \citep[e.g.,][]{Neeleman2017CiiFromDlaHost, Neeleman2019CiiEmissionFromHighzDLAs}, further away than similar galaxy-absorber pairs at low redshift. These results suggest that at high redshift, the cool CGM was particularly gas-rich.

Additional evidence comes from the diffuse \hi Lyman-$\alpha$ (\lya) emission that surrounds almost every star-forming galaxy at $z\gtrsim2$, with exponential scale lengths of \SIrange{1}{10}{\kilo\parsec} for individual systems \citep[\lya halos or LAHs; e.g.,][]{Wisotzki2016ExtendedLyaHalosMUSE, Leclerq2017MuseHudfLyaHaloes,Claeyssens2022LlamasPaper1} and up to \SI{100}{\kilo\parsec} for the so-called \lya ``blobs'' \citep[LABs; e.g.,][]{Francis1996LyaGroup, LeFevre1996ClusteringAroundRadioGalaxy, Steidel2000LyaProtoCluster, Venemans2002ProtoclusterRedshiftFour, Matsuda2004SubaruLyaBlobs}, which often host several galaxies, but not necessarily AGN \citep{Geach2009LabsPoweredByHeating}. Given the intrinsic complexities in interpreting the resonant, and highly dust-sensitive \lya line, the origins of extended \lya emission are still unclear. However, most of the proposed scenarios involve the presence of neutral hydrogen clouds in the CGM \citep[see][for a recent review]{Ouchi2020LymanAlphaReview}.

Moving away from the challenges of \lya, deep (sub)millimeter observations have revealed extended gas reservoirs of colder gas around high-$z$ galaxies in a variety of environments.
For example, several studies report extended CO and/or \ci emission around $z\approx2$ protocluster cores \citep{Emonts2014CoSurveyHighzRGs, Emonts2015CoRichMergerDragonfly, Emonts2016MolGasClusterHighZ, Emonts2018SpiderwebCircumgalacticCi, Emonts2019ColdCgmMammoth1, Ginolfi2017Molecular,Frayer2018MassiveMolecularComponentSmg, Li2021OutflowAndColdHaloQso, Umehata2021CiiObsOfLAB1, Cicone2021GiantMolecHaloRedshiftTwo}.
At even higher redshifts, the \ciim line becomes easily observable from the ground, and in the past decade it has proven very efficient at tracing extended gas in both extremely active \citep[i.e., quasars or starbursts, e.g.,][]{Carniani2013ExtendedCiiComplexMerger, Cicone2015ExtendedHaloQso, Ginolfi2020AlpineMergerTidal} and normal, less massive systems \citep{Fujimoto2019CiiHalos, Fujimoto2020AlpineHalos, HerreraCamus2021OutflowHZ4, Akins2022ExtendedCoolGasOutflowRedshift7, Lambert2023CiiHaloAroundHz7}. 
In particular, \citet{Fujimoto2020AlpineHalos}  finds that roughly 30\% of massive, isolated main sequence galaxies at $4<z<6$ display \SI{10}{\kilo\parsec}-scale ``\cii halos'', defined as the cases where significant \cii emission is detected at \SI{10}{\kilo\parsec} from the source while the UV and far infrared (FIR) emission are not.
Similar objects have been reported with deeper, higher angular resolution Atacama Large Millimeter/Submillimeter Array (ALMA) observations than those used by \citeauthor{Fujimoto2020AlpineHalos} \citep[e.g.,][]{HerreraCamus2021OutflowHZ4, Lambert2023CiiHaloAroundHz7}.
In all of these cases, however, the origin of extended \cii line emission remains unclear.

In dense regions, tidal interactions, cold accretion, outflows and AGN feedback seem to contribute to the presence of extended \cii line emission. 
In contrast, star-formation-driven outflows are often quoted as the most likely origin of extended \cii emission around individual and more isolated UV-bright galaxies \citep{Fujimoto2020AlpineHalos, HerreraCamus2021OutflowHZ4, Pizzati2020OutflowsCiiHaloes, Pizzati2023CiiHaloesSmokingGun}. However, confirming and understanding extended \cii emission at high redshift requires deeper and higher angular resolution ALMA observations.

CRISTAL stands for ``\cii Resolved ISm in STar-forming galaxies with ALma'' (Herrera-Camus et al. in prep.), and is an ALMA Cycle 8 Large Program that observed the \cii line and dust continuum emission of 19 main-sequence, star-forming galaxies at $4 < z < 6$ with a $\sim \ang{;;0.2}$ resolution.
CRISTAL builds on top of the highly successful ALPINE survey \citep{LeFevre2020AlpineSurveyDesign}, which conducted a wider census of \cii and dust at $\approx \ang{;;1}$ resolution in the COSMOS and GOODS-S fields.
Out of the 75 \cii-detected galaxies in the ALPINE sample \citep{Bethermin2020DataProcessing}, and based on the multiwavelength properties presented by \citet{Faisst2020AlpineMultiWavelength}, CRISTAL selected 19 sources that (1) have a specific SFR within a factor of 3 of the star-forming main sequence at their corresponding redshift; (2) have HST imaging available; and (3) have a stellar mass larger than $\log\left(M_\mathrm{star}/\mathrm{M}_\odot\right)\geq 9.5$. In addition, six sources from the ALMA archive that met the selection criteria were added to the sample, hence the total size of the sample is 25. The survey is designed to unveil detailed kinematics, search for resolved outflows, constrain ISM excitation and also probe extended emission.

Since the CRISTAL program targets the massive end of its parent sample, it is not surprising that many of them show clear signs of multiplicity or interaction (e.g., Ikeda et al. in prep, Lee et al. in prep., Posses et al. in prep.). Among them, CRISTAL-01 stands out due to its proximity ($\sim$\ang{;;1.6}) to the well-known sub-millimeter galaxy \object{AzTEC\,J100055.19+023432.8} at $z=4.54$ \citep[\sysname;][]{Capak2008ExtremeStarburst, Aretxaga2011AztecSurveyCosmos}. Here, we present new ALMA observations of this system and report the discovery of a puzzling \cii-emitting gas plume that extends from the center of the system. 

The paper is organized as follows: in Sec.~\ref{sec:backstory} we give an overview of the literature on this particular system. Sec.~\ref{sec:data} describes the observations and reduction of the new ALMA dataset and the ancillary VLT/MUSE and HST data.
In Sec.~\ref{sec:results} we detail the analysis steps and present the results characterizing the \cii plume. 
Next, in Sec.~\ref{sec:discussion} we explore different physical scenarios that could give rise to the observed emission. 
Finally, Sec.~\ref{sec:conclusions} closes with a summary and the main conclusions.

Throughout the paper, we adopt a flat $\Lambda$CDM cosmology with $H_0=\SI{70}{\kilo\meter\per\second\per\mega\parsec}$ and $\Omega_{m,0}=0.3$.
Under this assumption, \SI{1}{\arcsecond} corresponds to \SI{6.57}{\kilo\parsec} of proper physical scale at $z=4.54$. When relevant, we adopted a \citet{Chabrier2003IMF} initial stellar mass function (IMF). 

\section{The J1000+0234 system}\label{sec:backstory}

\begin{table}[!htb]
  \centering
  \caption{Global properties of the two main galaxies in the \sysname system.}\label{tab:glob_tab}
  \renewcommand{\arraystretch}{1.5}
  \begin{tabular}{c|cc|cc}
	  \hline
	  \multirow{2}{*}{Property}  & \multicolumn{2}{c|}{\sysname-North} &  \multicolumn{2}{c}{CRISTAL-01a} \\
				      \cline{2-5} 
	            & Value & Ref. & Value & Ref. \\
		    \hline
	  \cii redshift	    & 4.5391 & 1 & 4.5537 & 2 \\
	  $\log\left(M_\mathrm{stars}\right)\,[\si{\msun}]$ & \num{10.14+-0.08} & 3 & \num{9.16+-0.07} & 3 \\
	  $\log\left(M_\mathrm{dyn}\right)\,[\si{\msun}]$ & \num{11.15+-0.19} & 1 & - & - \\
		  UV slope $(\beta_\mathrm{UV})$ & $-1.01_{-0.32}^{+0.39}$ & 3 & $-2.04_{-0.11}^{+0.12}$ &  3\\
		  SFR$_\mathrm{UV}$ [\si{\msun\per\year}] & \num{52.6+-8.5} & 3 & \num{147.6+-7.4} & 3 \\
		  SFR$_\mathrm{IR}$ [\si{\msun\per\year}] & $440_{-320}^{+1200}$ & 3 & \num{56+-35}\tablefootmark{a} & - \\
		  SFR$_\mathrm{tot}$ [\si{\msun\per\year}] & $490_{-320}^{+1200}$ & 3 & \num{204+-35} & - \\
		  \hline
		  \hline
  \end{tabular}
  \tablebib{(1)~\citet{Fraternali2021FastRotators}; (2)~\citet{Bethermin2020DataProcessing}; (3)~\citetalias{GomezGuijarro2018AlmaMinorMergersSmgs}.}
  \tablefoot{
  \tablefoottext{a}{Estimated from the IRX-$\beta_\mathrm{UV}$ relation \citep{Meurer1999IrxBetaRelation} assuming an SMC attenuation law \citep{Bouchet1985SmcExtinctionLaw}.}}
\end{table}

In this paper, we study the core region of the \sysname system at $z=4.54$, first reported by \citet[][]{Capak2008ExtremeStarburst} as a bright sub-millimeter source with an associated \lya blob.
This region (see Fig.~\ref{fig:maps}) hosts two highly star-forming  galaxies within $\approx \SI{20}{\kilo\parsec}$ in projection \citep[][hereafter GG18]{GomezGuijarro2018AlmaMinorMergersSmgs}. It is embedded in a larger scale overdensity of galaxies \citep{Smolcic2017SmgEnvironments, Loiacono2021AlpineCiiLf}, and possibly linked to the PCI\,J1001+0220 protocluster \citep{Lemaux2018VudsProtoCluster}.
One of the central galaxies, previously known as J1000+0234-South \citepalias{GomezGuijarro2018AlmaMinorMergersSmgs} but hereafter called \ca, contributes $\approx$75\% of the total the rest-frame UV emission of the pair, and belongs to the sample of Lyman-break galaxies (LBGs) targeted by both the ALPINE and CRISTAL surveys. 
The other galaxy, J1000+0234-North, lies merely \ang{;;1.6} from \ca and accounts for all of the observed submillimeter flux ($S_{\SI{870}{\micro\meter}} = \SI{7.8+-0.2}{\milli\jansky}$; \citetalias{GomezGuijarro2018AlmaMinorMergersSmgs}) and most of the stellar mass ($\approx \SI{2e10}{\msun}$; e.g., \citealt{Schinnerer2008MolGasMajorMerger}; \citetalias{GomezGuijarro2018AlmaMinorMergersSmgs}).
Hereafter, we will also refer to it as ``the DSFG'' (dusty star-forming galaxy).
The global properties of each galaxy are extracted from the literature and listed in Table~\ref{tab:glob_tab}.

Multiple follow-up studies have targeted \sysname as one of the brightest and most extreme non-quasar systems known at $z\gtrsim4$.
Until recently, the general picture depcits \ca and the DSFG undergoing a merger event, which potentially drives the elevated SFRs (\citealt{Capak2008ExtremeStarburst, Schinnerer2008MolGasMajorMerger, Smolcic2015PhysicalPropertiesSmgsCosmos}; \citetalias{GomezGuijarro2018AlmaMinorMergersSmgs}).
Yet ALMA observations of the \cii line later revealed that \cb rotates fast at $V_\mathrm{rot}\approx\SI{500}{\kilo\meter\per\second}$ with $V/\sigma \gtrsim 9$, suggesting a dynamically cold gas component \citep{Jones2017DynamicsTiltedRingSmgs, Fraternali2021FastRotators}. 
In other words, the merger is either not massive enough or has not had time to dynamically disrupt the internal kinematics of \cb. This is consistent with SED models that put the stellar mass ratio between \ca and \cb at $\sim$1:10 \citepalias{GomezGuijarro2018AlmaMinorMergersSmgs}.

Current Chandra pointings do not detect X-rays from \sysname, putting an upper limit on luminosity of about \SI{1.3e43}{\erg\per\second} in the 0.5-2 keV band \citep{Smolcic2015PhysicalPropertiesSmgsCosmos}. This value is considerably higher than the \SI{1e42}{\erg\per\second} traditional threshold for AGN identification \citep[e.g.,][]{Szokoly2004Cdfs}.
Moreover \cb's radio emission is weak \citep{Carilli2008RadioLbgs}, and consistent with the infrared-radio correlation \citep{Smolcic2015PhysicalPropertiesSmgsCosmos}. However, \citet[][hereafter J23]{Jimenez-Andrade2023LyaNebulaeGalaxyPair} recently obtained VLT/MUSE observations of \sysname, that yielded not only a very high fidelity 3D IFU map of the LAB in which \sysname is embedded, but also the detection of spatially extended \civ and \heii emission. The authors argue that the high \civ/\lya and \heii/\lya  ratios can be explained by the presence of an AGN in \cb. While this claim still needs confirmation, \citetalias{Jimenez-Andrade2023LyaNebulaeGalaxyPair} provided a first view of the complex CGM of \sysname, as well as spectroscopic evidence for the overdensity, after identifying five \lya emitters within the MUSE field of view ($1\times1\,\mathrm{arcmin}^2$).

In parallel to the studies focusing on the massive DSFG, \ca was independently targeted by the ALMA ALPINE survey of \cii emission in bright LBGs \citep[][therein labeled as DEIMOS-COSMOS 842313]{LeFevre2020AlpineSurveyDesign}. A successful detection of the line provided the first systemic redshift of this source at $z=4.5537$ \citep{Bethermin2020DataProcessing}, closely matching the previously reported \lya redshift \citep[$z=\num{4.5520}$;][]{Hasinger2018DeimosCosmos}. In the next section, we describe the high-angular resolution ALMA observations obtained by CRISTAL, along with the rest of the observations used in this article.

\section{Observations \& data reduction}\label{sec:data}
    \subsection{ALMA}\label{sec:data:alma}
    We use \textsc{casa} \citep[version 6.5.2,][]{CasaTeam2022CasaRefPaper} to combine ALMA observations of the redshifted \ciim line targeting \sysname from three different programs, namely the ALPINE survey \citep{LeFevre2020AlpineSurveyDesign}, the CRISTAL survey (Herrera-Camus et al. in prep.), and an archival dataset from project 2019.1.01587.S (PI: F. Lelli). The ALPINE observations were carried out with the most compact configuration (C43-1), tuned to cover the line at \SI{349.1}{\giga\hertz} with a velocity resolution of \SI{40}{\kilo\meter\per\second} and a natural weighting beam size of $\SI{1.25}{\arcsecond}\times\SI{0.78}{\arcsecond}$ \citep{Bethermin2020DataProcessing}. The CRISTAL data, on the other hand, include deeper integrations in two antenna configurations, namely C43-1 and C43-4, and with a higher spectral resolution (\SI{10}{\kilo\meter\per\second} per channel) than that of the ALPINE data. These observations were designed to resolve the \cii emission with a beam of $\sim$\SI{0.25}{\arcsecond}, equivalent to \SI{1.65}{\kilo\parsec} at $z = 4.54$. Finally, the 2019.1.01587.S dataset was observed using a longer baseline configuration (C43-6), providing a nominal resolution of \ang{;;0.06}. However, the spectral windows were tuned around the DSFG's rest-frame velocity, so the frequency overlap with previous data is partial and only covers the red half of \ca's emission line. 
    We do not include data from the Cycle 2 program 2012.1.00978.S \citep[PI: A. Karim;][]{Jones2017DynamicsTiltedRingSmgs, Fraternali2021FastRotators} into the combined dataset, because its shallower depth plus the complexities of weighting data that was processed with old versions of the pipeline\footnote{\url{https://casaguides.nrao.edu/index.php/DataWeightsAndCombination}} would have resulted in a marginal improvement.

    In addition, thanks to the brightness of \cb  we performed self-calibration on the continuum visibilities of both ALPINE and CRISTAL datasets before combination. This was done in two ``phase-only'' rounds for each observation, the first one combining spectral windows and scans, and the second one only the scans (of average length $\approx\SI{180}{\second}$). This process resulted in a $\approx$11\% decrease in continuum rms and the mitigation of patchy patterns in the noise.

    Finally, the self-calibrated and combined measurement sets were processed with CRISTAL's reduction pipeline as described in Herrera-Camus et al. (in prep.). Briefly, it starts by subtracting the continuum on the visibility space using \textsc{casa}'s \verb|uvcontsub| task. After that, it runs \verb|tclean| with automasking multiple times, producing cubes with different weightings and channel widths. In all cases the data are cleaned down to $1\sigma$. In this paper, we use datacubes with \SI{20}{\kilo\meter\per\second} channel width and either natural or Briggs (\verb|robust=0.5|) weighting.

    Since the data combines different array configurations, it is important to measure and apply the ``JvM'' correction \citep{JorsaterAndvonMoorsel1995BarredSpiralHI}. We do this by following the method of \citet{Czekala2021JvMGuidelines}. This correction takes into account the significant deviations from Gaussianity that the core  of the dirty beam can have in multi-array observations and ensures that both the convolved CLEAN model and the residuals have compatible units. While the CRISTAL pipeline provides JvM-corrected products, it uses a single correction factor per spectral window. Since we combine multiple datasets from different projects, the uv coverage has significant variations within a spectral window. For this reason, we compute and apply the correction in channel ranges with similar beam properties. We find a mean multiplicative correction factor of $\epsilon\approx 0.37$ in the channels covering the \cii line emission when using Briggs weighting, and $\epsilon\approx0.31$ when using natural weighting.

    \subsection{MUSE}\label{sec:data:muse}
    We retrieved observations of the \sysname field from the ESO archive taken with the Multi Unit Spectroscopic Explorer (MUSE) instrument mounted on the Very Large Telescope (UT4-Yepun) using ground-layer adaptive optics and the Wide Field Mode. 
    These observations, comprising 16 exposures of \SI{900}{\second} (total \SI{4}{\hour}), were taken as part of ESO GTO programs 0102.A-0448 (PI: S. Lilly) and 0103.A-0272 (PI: S. Cantalupo) under good weather conditions  with average seeing of \SI{0.9}{\arcsecond} and airmass below 1.4. Here we use an independent reduction from that of \citetalias{Jimenez-Andrade2023LyaNebulaeGalaxyPair}, but we refer the reader to their work for further details about the observations.

The standard calibrations and procedures are performed using the MUSE pipeline \citep[version 2.8.3;][]{Weilbacher2020MUSE_DPP} within the ESO Recipe Execution Tool (EsoRex) environment \citep{ESO2015CPL}. The wavelength solution is set to vacuum. We then apply the Zurich Atmosphere Purge \citep[ZAP, version 2.1;][]{Soto2016ZAP} post-processing tool to further remove sky line residuals. As described by \citetalias{Jimenez-Andrade2023LyaNebulaeGalaxyPair}, half of the exposures were affected by intra-dome light contamination, resulting in excess counts between \SI{8000}{\angstrom} and \SI{9000}{\angstrom}. We decided to use all the exposures in the combined datacube regardless, since the issue does not affect wavelengths near the \lya emission at \SI{6742}{\angstrom}. 
   We then follow the steps described in \citet{Solimano2022LymanAlphaSGASJ1226} to scale the variance cube and match the observed noise levels.
   Finally, we apply a simple 2D translation to the WCS to match the positions of the two Gaia DR2 \citep{GaiaCollaboration2018Release2} sources in the field.
   We check the alignment of the cube with respect to HST by producing pseudo broadband image from the ACS/F814W filter curve. Matching sources in the MUSE pseudo-F814W image and the ACS/F814W image yield an astrometric rms of 220 mas, or about one MUSE pixel.

    The resulting datacube  yields a $1\sigma$ noise level of $\approx$\SI{1e-19}{\erg\per\second\per\centi\meter\squared\per\arcsec\squared} per spectral layer around \SI{6750}{\angstrom}, computed from randomly placed apertures of \SI{1}{\arcsec\squared} area. 
    The datacube is sampled at pixel scale \ang{;;0.2} and spectral layers have a width of \SI{1.25}{\angstrom}.
    We fit a Moffat profile to an $r=19\,\mathrm{mag}$ G-type star in the field and find a point spread function (PSF) FWHM of \ang{;;0.6} around \SI{6750}{\angstrom}.
    At this wavelength, the instrument yields a resolving power of $R=\num{2538}$. 

    \subsection{HST}\label{sec:data:hst}
    We retrieved from MAST\footnote{\url{https://mast.stsci.edu/portal/Mashup/Clients/Mast/Portal.html}} all the available HST data for the \sysname field.
    We found observations in the ACS/F606W, ACS/F814W, WFC3/F105W, WFC3/F125W and WFC3/F160W bands, covering \SIrange{1000}{3200}{\angstrom} in the rest-frame at $z=4.54$.
    Images were processed using the standard pipeline, co-added and aligned to Gaia DR2 \citep{GaiaCollaboration2018Release2}. All images were then drizzled  with a square kernel and a pixel fraction of \num{0.5} using the \textsc{astrodrizzle} routine from \textsc{DrizzlePac} \citep{Stsci2012DrizzlePac, Hack2021DrizzlePacv3}, executed within the \textsc{grizli} pipeline \citep{Brammer2021GrizliSoftware}. ACS images were drizzled to a common pixel size of \ang{;;0.03} while for WFC3/IR images we used a pixel size of \ang{;;0.06}.

    \begin{figure*}[!htb]
        \centering
        \includegraphics[width=17cm]{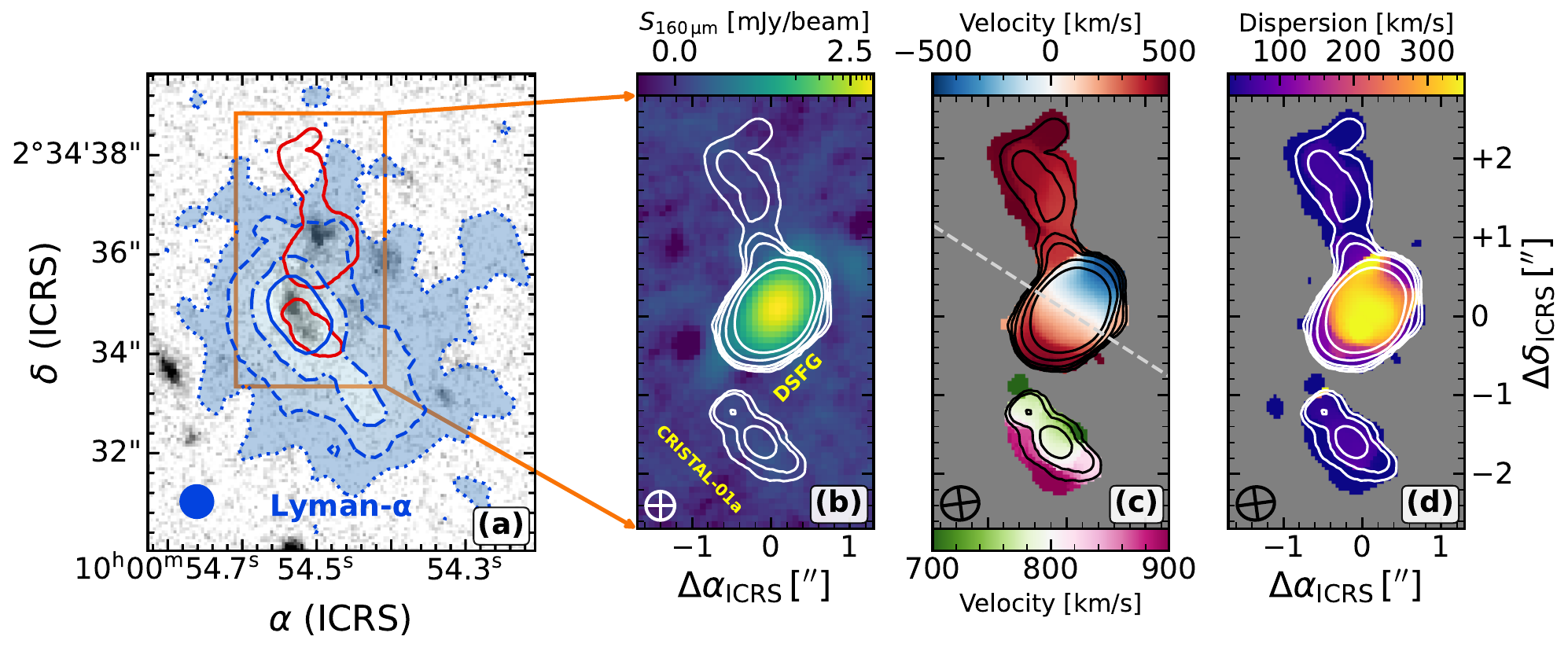}
	\caption{HST, \lya, and \cii morphologies.
	\textbf{(a).} $\ang{;;10}\times\ang{;;6}$ cutout of the WFC3/F160W image in grayscale. The blue-filled contours represent $\left\lbrace1, 5, 16, 30\right\rbrace \times 10^{-18}\,\si{\erg\per\second\per\centi\meter\squared\per\arcsec\squared}$ levels of \lya surface brightness based on the adaptive narrowband image described in the main text. The diameter of the blue circle represents the FWHM$=\ang{;;0.67}$ of the PSF of the VLT/MUSE observations. The orange box outlines the $\ang{;;2.5}\times \ang{;;5}$ zoom-in region displayed in the next panels.
	\textbf{(b).} ALMA Band 7 continuum image in logarithmic stretch. The white contours follow \cii emission at $\left\lbrace0.01, 0.03, 0.07, 0.2, 0.5\right\rbrace \times \si{\jansky\kilo\meter\per\second\per\beam}$, and highlight the non-detection of dust continuum in either \ca or the plume. 
\textbf{(c)}. Adaptively-masked \cii velocity field (moment-1). A single colormap is assigned separately to \cb and \ca. The midpoint of the colormaps matches their corresponding systemic velocity, with the zero set at the redshift of \cb, $z_\mathrm{[C II]}=4.5391$. Again, overlaid contours show increasing levels of \cii SB. The dashed line indicates the projected rotation axis at a $\mathrm{PA}=\ang{57.4}$ through the \cii kinematic center of \cb (see Appendix~\ref{sec:appendixA}).
		\textbf{(d)}. Adaptively-masked \cii velocity dispersion map (moment-2), with \cii SB contours.
	}
        \label{fig:maps}
    \end{figure*}

\section{Results \& Analysis}\label{sec:results}
    \subsection{Adaptive masking of datacubes}\label{sec:results:masking}
    Extracting the total flux and spatial extent of the different types of emission considered here, requires taking into account the contributions of faint and diffuse components. Now, given the complexity and spatial variations of the \lya and \cii profiles in the \sysname system, a simple pseudo-narrowband ``collapse'' around the line will hide narrow and low surface brightness (SB)  features below the noise level. Instead, we adopt an adaptive approach based on the ``matched filtering'' technique. This involves creating a 3D mask that takes into account the line morphology. For the MUSE data, we used the off-the-shelf software \textsc{LSDcat} \citep{Herenz2017LsdCat}, while for the ALMA cubes we used a custom script.

   \textsc{LSDcat} works by building an optimized S/N detection derived from the cross-correlation of the data with a template signal. Here, we first removed the continuum emission by running a median filter over the spectral axis across the full wavelength range of the cube, using the default window width of 151 spectral pixels (\SI{188.75}{\angstrom}). This step yields a continuum-only datacube which is then subtracted from the original cube. The next step is optimized for our blind search of \lya emitters in the cube, but it also performs well on enhancing the low SB features. The results of our search and the subsequent characterization of the detected objects will be presented in a separate paper.

   Following \citet{Herenz2017LsdCat} we built our template as a point-like source with a Gaussian spectral profile of $\mathrm{FWHM}=\SI{250}{\kilo\meter\per\second}$, a choice that maximizes sensitivity to faint and compact line emitters. After convolving the continuum-subtracted cube by the PSF (spatial filter), we convolved the resulting cube with a \SI{250}{\kilo\meter\per\second} Gaussian kernel (spectral filter) to construct the 3D matched filter output. We took into account the wavelength dependence of the PSF by fitting 2D Moffat profiles to an isolated bright star in the original cube, using 20 wavelength bins. We simultaneously fit the Moffat parameters' dependence on wavelength using  3\textsuperscript{rd} and 2\textsuperscript{nd} order polynomials, for the FWHM and power index parameters, respectively. These polynomials were then used to interpolate the PSF to all the channels of the cube. Then, we computed the detection S/N cube as the voxel-by-voxel ratio between the filtered datacube and the square root of the propagated variances.

    Finally, we selected all voxels with $\mathrm{S/N}\geq 2$ between \SI{6729}{\angstrom} and \SI{6776}{\angstrom} (equivalent to a velocity range of [\SI{-947}{\kilo\meter\per\second}, \SI{1135}{\kilo\meter\per\second}]). To refine the selection, we exclude spaxels in which less than three voxels are above the S/N threshold. Furthermore, we create a 2D mask of the spaxels that satisfy these conditions and subsequently prune the regions with less than 55 connected spaxels. We find that this number successfully masks any remaining spurious signal. After we remove the corresponding voxels, we apply the resulting 3D mask to the data, and integrate along the wavelength axis. The result is shown in Fig.~\ref{fig:maps}a as filled contours overlaid on top of an HST image. We recover an irregular and extended \lya morphology down to \SI{1e-18}{\erg\per\second\per\centi\meter\squared\per\arcsec\squared}, in agreement with \citetalias{Jimenez-Andrade2023LyaNebulaeGalaxyPair}. The highest SB emission is centered on \ca, but extends to a secondary peak $\sim\ang{;;1.5}$ to the southwest.

    We applied a similar procedure to the \cii cube.  Starting from the naturally-weighted cube binned in \SI{20}{\kilo\meter\per\second} channels, we convolved with a $\sigma=\SI{30}{\kilo\meter\per\second}$ Gaussian kernel along the velocity axis, and a $\sigma=\ang{;;0.1}$ 2D Gaussian kernel in the spatial axes. We then measured the rms in the signal-free regions of the convolved cube. Finally, we split cells above and below a $2\times\mathrm{rms}$ threshold into a 3D mask, which we then fed to \textsc{casa} task \verb|immoments| to obtain the intensity, velocity, and velocity dispersion maps from the original cube (as shown from panels b to d in Fig.~\ref{fig:maps}). 

This approach recovers the bright \cii emission from \cb but also reveals faint and extended emission in \ca, and most notably, in a \ang{;;2.4}-long plume extending north of \cb. We note, however, that this elongated diffuse emission was already apparent in the individual channels of our cubes, as will be discussed in Sec.~\ref{sec:results:cii_spectral}. 
The intensity map spans more than two orders of magnitude in SB, from \SI{0.01}{\jansky\kilo\meter\per\second\per\beam} (outermost contour in Fig.~\ref{fig:maps}b-d) to \SI{2.5}{\jansky\kilo\meter\per\second\per\beam} at the center of the DSFG, equivalent to \SI{5e5}{\lumsol\per\kilo\parsec\squared} and \SI{1.25e8}{\lumsol\per\kilo\parsec\squared} respectively at $z=4.54$. If we take the \cii SB as a tracer of SFR surface density, according to the local $\Sigma_\mathrm{[CII]}-\Sigma_\mathrm{SFR}$ relation \citep{HerreraCamus2015CiiAsSfrTracerKingFish}, the lower limit  would correspond to $\Sigma_\mathrm{SFR}=\SI{0.01}{\msun\per\year\per\kilo\parsec\squared}$, which is well within the regime of normal star-forming galaxies in the Local Universe from the KINGFISH sample \citep{Kennicutt2011KingfishSample}.

    In the velocity space, significant \cii emission spans from \SI{-500}{\kilo\meter\per\second} to \SI{850}{\kilo\meter\per\second} across smooth gradients (see Fig.~\ref{fig:maps}c). The main gradient goes along the major axis of the DSFG, as previously found by \citet{Jones2017DynamicsTiltedRingSmgs} and \citet{Fraternali2021FastRotators}. Interestingly, the plume displays a gradient along its long axis, with a mild increase in velocity as one moves away from the DSFG. While the plume meets the DSFG in the approaching side, the velocity of the plume overlaps with that of the receding side of the DSFG. Finally, \ca's \cii emission is centered at $v\approx\SI{800}{\kilo\meter\per\second}$, and shows a more irregular gradient approximately aligned with the minor axis.
 
    The velocity dispersion (Fig.~\ref{fig:maps}d), on the other hand, is largest at the center of the DSFG, with the other structures showing low values ($\sigma \lesssim \SI{100}{\kilo\meter\per\second}$) and little to no variation.

    \subsection{Parametric morphology}\label{sec:results:morpho}
    In this section we study the morphological parameters of the two galaxies, \ca and \cb, in the \sysname system.
    To this end, we make use of 2D light profile modeling code \textsc{PyAutoGalaxy} \citep{Nightingale2023PyAutoGalaxy} built on top of the \textsc{PyAutoFit} \citep{Nightingale2021PyAutoFit} probabilistic programming framework.
    While \textsc{PyAutoGalaxy} is capable of directly fitting the interferometric visibilities, in this paper we used the image-based fitter for a faster workflow. 
    We account for the noise correlation in the images by feeding the full covariance matrix into the calculation of the likelihood.
    The covariance matrix is estimated in source-free regions of the image using the method and the code\footnote{ESSENCE, \citealt{Tsukui2023Essence}} presented by \citet{Tsukui2022CorrelatedNoiseInterferometry}.

    We start by modeling \cb, which shows a regular and almost symmetric shape. For this reason, we choose to fit a single 2D S\'ersic profile \citep{Sersic1968AtlasdeGalaxias}, with a total of seven free parameters. 

    We performed two independent fits. One for the rest-frame \SI{160}{\micro\meter} continuum map shown in panel b of Fig.~\ref{fig:maps}, and the other for the \cii integrated intensity map.  However, we refrained from using the adaptive intensity map, since the noise properties are not well defined after the masking procedure. Instead, we used regular, unmasked intensity maps integrated within a given velocity range. For \cb, we integrated between \SI{342.293}{\giga\hertz} and \SI{344.006}{\giga\hertz}, corresponding to a bandwidth of \SI{1492.7}{\kilo\meter\per\second}. For the center coordinates and the effective radius, we adopted broad Gaussian priors centered on previously published values \citep{Fraternali2021FastRotators}. For the intensity parameter we adopted a uniform prior between 0 and the maximum surface brightness of the image. Finally, based on previous work that characterized this source as a disk \citep[e.g.][]{Jones2017DynamicsTiltedRingSmgs}, we adopt a Gaussian prior for the S\'ersic index centered at $n=1$ (exponential disk) with a $\sigma=1$, but allowed $n$ to vary between 0.2 and 10.

    The parameter space was explored using the \textsc{dynesty} \citep{Speagle2020Dynesty, Koposov2022DynestyZenodoV203} nested sampler backend with 50 live points. Table~\ref{tab:dsfg_param} lists the results of fitting with the parameters values and uncertainties drawn from the Bayesian posterior probability distribution.

    \begin{table}[!hbt]
	    \centering
	    \caption{Results of the parametric 2D fitting of exponential profiles to the \SI{160}{\micro\meter} continuum and \cii line maps of \cb.}\label{tab:dsfg_param}
	    \begin{tabular}{lc}
		    \hline
		    Property & Value \\
		    \hline
		    \multicolumn{2}{c}{\it \SI{160}{\micro\meter} continuum} \\
		    Center (RA) & $10^\mathrm{h}00^\mathrm{m}54.49129^\mathrm{s}\pm0.00002^\mathrm{s}$\\
		    Center (Dec.) &  $\ang{+02;34;36.120}\,\pm\,\ang{;;0.0002}$ \\
		    $R_\mathrm{eff}$ [pkpc] & \num{0.740+-0.003} \\
		    S\'ersic index ($n$) & \num{1.29+-0.01}\\
		    Axis ratio (minor/major) & \num{0.400+-0.002} \\
		    Flux density [\si{\milli\jansky}] & \num{8.03+-0.03} \\
		    PA [degrees] & \num{57.8+-0.1} \\

		    \multicolumn{2}{c}{\it \cii emisison}\\

		    Center (RA) & $10^\mathrm{h}00^\mathrm{m}54.4904^\mathrm{s}\pm0.0003^\mathrm{s}$\\
		    Center (Dec.) & $\ang{+02;34;36.140}\,\pm\,\ang{;;0.005}$ \\
		    $R_\mathrm{eff}$ [pkpc] & \num{1.13+-0.04} \\
		    S\'ersic index ($n$) & \num{0.71+-0.08}\\
		    Axis ratio (minor/major) & \num{0.34+-0.01} \\
		    Integrated flux [\si{\jansky\kilo\meter\per\second}] & \num{7.8+-0.3} \\
		    PA [degrees] &  \num{57.4+-0.8} \\
		    \hline
		    \hline
    	\end{tabular}
	\tablefoot{Radii are circularized and given in physical kiloparsecs (pkpc).}
    \end{table}

\begin{figure}[!hbt]
    	\centering
	\resizebox{\hsize}{!}{\includegraphics{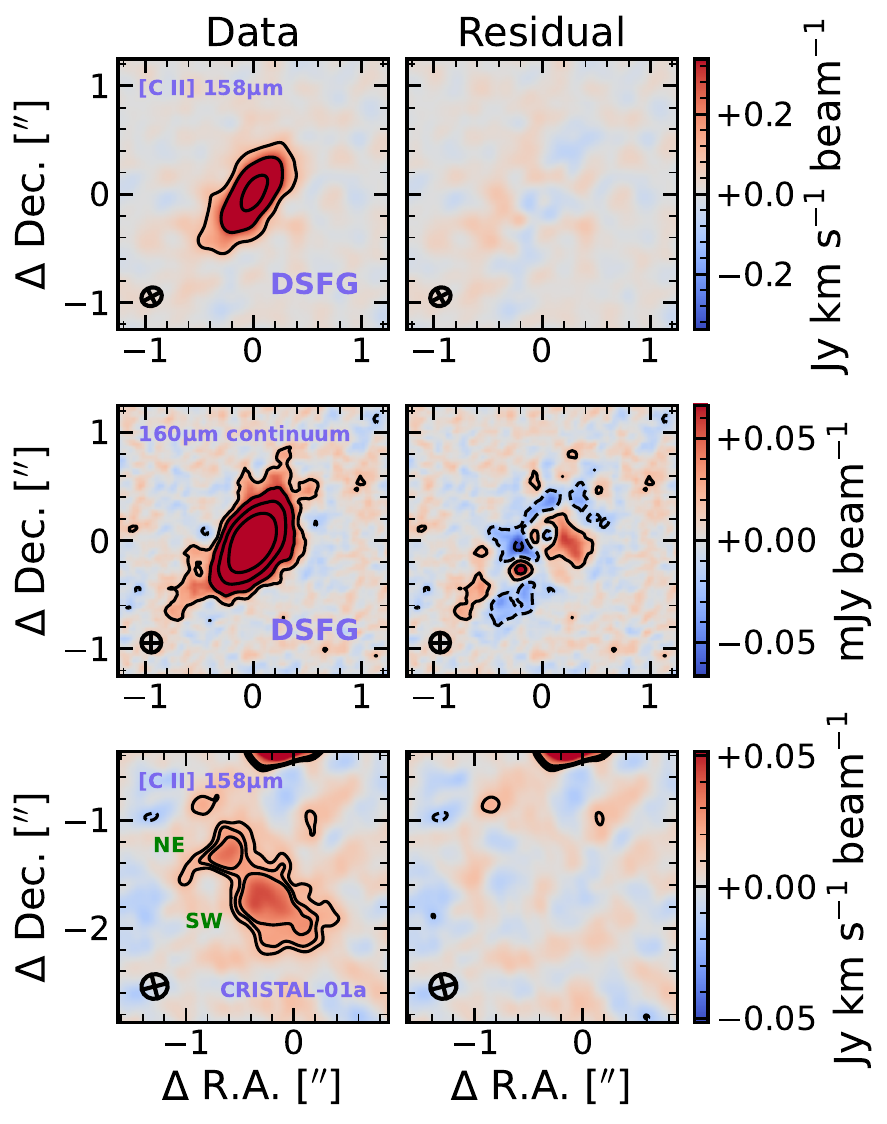}}
	\caption{Results from parametric 2D modeling with \textsc{PyAutoGalaxy}. Each row shows a different source and/or image. Left column displays the observed emission, while the right column shows the residuals after subtracting the maximum likelihood model. The first and second rows show the modeling of the DSFG's \cii emission and \SI{160}{\micro\meter} dust continuum, respectively. Black contours represent the $\pm3,9,27$ and $81\sigma$ levels. Results for the \cii emission of \ca (CRISTAL-1) are shown in the third row. Here, the contours only trace $\pm3,4$ and $5\sigma$ levels. }
	\label{fig:residuals}
    \end{figure}

    Our analysis reveals that the \cii and dust morphologies are slightly different in \cb. The effective radius of the gas component is approximately 1.5 times larger than that of the dust, while the axis ratio is smaller. However, since the S\'ersic indices and the S/N are not equal between \cii line and \SI{160}{\micro\meter} continuum, we refrain from claiming the gas is more extended than the dust. Assuming instead that the dust and gas  both have disk-like geometries, share the same inclination, and are optically thin, the difference in axis ratio suggests that the dust component is thicker in the polar direction relative to the gas component. The higher S\'ersic index also indicates that dust is more centrally concentrated, yet the residual continuum map shows significant features that extend beyond the central region and do not follow axial symmetry (see middle row of Fig.~\ref{fig:residuals}).

    Next, we move to \ca and fit only the \cii emission, since the continuum was not detected. We construct the unmasked intensity map by integrating the cube between \SI{342.088}{\giga\hertz} and \SI{342.454}{\giga\hertz}, which encloses only 85\% of the total line flux but maximizes the S/N of the map. The source clearly breaks into two subcomponents, resembling the rest-frame UV morphology, albeit offset $\approx$\ang{;;0.3} to the Southwest. We thus model the emission with two independent profiles, an elliptical 2D S\'ersic for the brighter component (hereafter SW clump) and a circular Gaussian profile for the fainter component (hereafter NE clump). This model has a total of 11 free parameters. For the SW clump we use the same prior on the S\'ersic index as before. For the rest of the quantities we use either broad uniform or Gaussian priors covering reasonable limits in the parameter space.

    The best-fit parameters and uncertainties from this fit are listed in Table~\ref{tab:c1a_param}. The right panel of Fig.~\ref{fig:residuals} shows the 2D residuals. 
    Notably, the SW clump alone has twice the effective radius of the DSFG's \cii emission and more than three times the radius of its dust emission. 
    This size is comparable to previous measurements performed on the F814W imaging \citep{Fujimoto2020AlpineHalos}. 
    If we were to associate the \cii clumps to the two UV clumps seen by HST, we notice the brightness order is inverted.
    The SW clump's \cii line flux density is about ten times that of than the NE clump. Yet in the UV, the NE clump is brighter. This discrepancy could be attributed to differences in dust attenuation. In fact, \citetalias{GomezGuijarro2018AlmaMinorMergersSmgs} provided maps of the rest-frame UV $\beta_\mathrm{UV}$ slope based on the HST imaging, and they show that the southern part of \ca is slightly redder ($\beta_\mathrm{UV}\approx-2.0$) than the northern part ($\beta_\mathrm{UV}\approx-2.2$).

\begin{table}[!htb]
  \centering
\caption{Results of the parametric 2D fitting to the \cii line map of \ca.}\label{tab:c1a_param}
  \begin{tabular}{lc}
    \hline
    Property & Value \\
    \hline
    \multicolumn{2}{c}{\it SW clump} \\
    Center (RA) & $10^\mathrm{h}00^\mathrm{m}54.505^\mathrm{s}\pm0.003^\mathrm{s}$\\
    Center (Dec.) &  $\ang{+02;34;34.34}\,\pm\,\ang{;;0.02}$ \\
    $R_\mathrm{eff}$ [pkpc] & \num{2.4+-0.4} \\
    S\'ersic index ($n$) & \num{1.0+-0.3}\\
    Axis ratio (minor/major) & \num{0.82+-0.09} \\
    Integrated flux [\si{\jansky\kilo\meter\per\second}] & \num{0.39+-0.08} \\
    PA [degrees] & \num{175+-45} \\
    \multicolumn{2}{c}{\it NE clump}  \\
    Center (RA) & $10^\mathrm{h}00^\mathrm{m}54.537^\mathrm{s}\pm0.02^\mathrm{s}$\\
    Center (Dec.) & $\ang{+02;34;34.75}\,\pm\,\ang{;;0.03}$ \\
    $R_\mathrm{eff}$ [pkpc] & \num{0.8+-0.3} \\
    Integrated flux [\si{\jansky\kilo\meter\per\second}] & \num{0.03+-0.02} \\
    \hline
    Total flux\tablefootmark{a} (SW + NE)[\si{\jansky\kilo\meter\per\second}] & \num{0.5+-0.2} \\
    Centroid separation [arcsec] & \num{0.63+-0.06} \\
    \hline
    \hline
    \end{tabular}
    \tablefoot{The table separates the fitted parameters for each clump. Radii are circularized and given in physical kiloparsec (pkpc). \tablefoottext{a}{Flux has been corrected by a factor 1.17 to account for emission outside the velocity integration range.}}
\end{table}

    \subsection{Spectral properties of the [C II]-emitting plume}\label{sec:results:cii_spectral}
\begin{figure*}[!hbt]
	\centering
	\includegraphics[width=17cm]{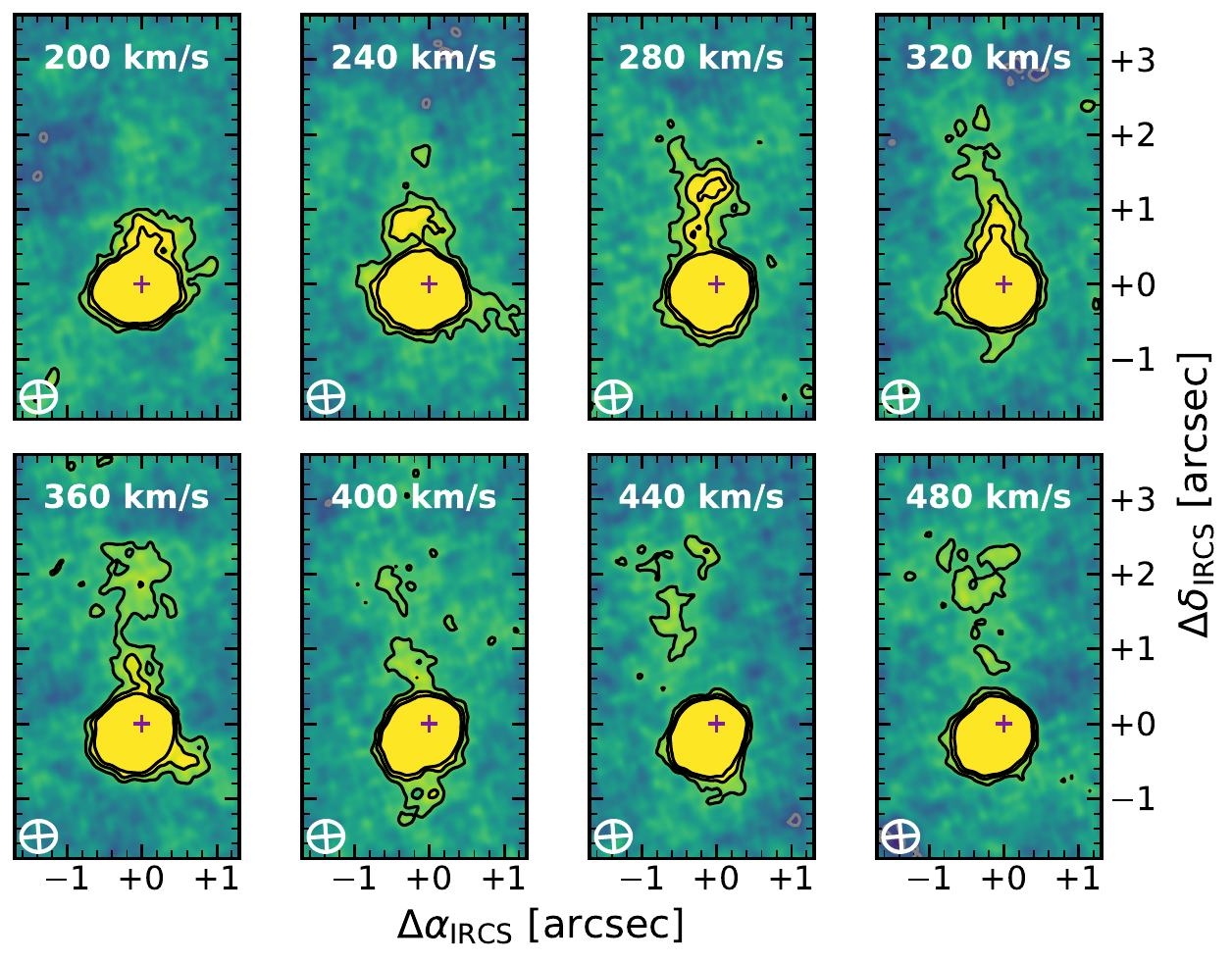}
	\caption{Selected channel maps from the low-resolution \cii datacube binned to \SI{40}{\kilo\meter\per\second}. Black contours represent 3, 5 and 7 times the JvM-corrected noise rms level, while the gray contours show the negative $3\times\mathrm{rms}$ level. The purple cross in each channel indicates the origin of the coordinates, at $\alpha_\mathrm{ICRS}=\ang{150.227063}$, $\delta_\mathrm{ICRS}=\ang{2.576679}$.}\label{fig:channel-maps}
\end{figure*}

    While the adaptive masking scheme described in Sect.~\ref{sec:results:masking} helped us identify all the \cii signal present in the cube and unveil the general kinematic trends, the noise properties of the masked maps in Fig.~\ref{fig:maps} remain undefined. Moreover, since our data display a large dynamic range in both surface brightness and velocity dispersion, the details of the plume are partly outshined by the DSFG. In order to get a clearer picture of the plume, Fig.~\ref{fig:channel-maps} displays eight consecutive channel maps of \SI{40}{\kilo\meter\per\second} width, covering from \SI{200}{\kilo\meter\per\second} to \SI{480}{\kilo\meter\per\second} relative to the systemic velocity of \cb.

The limits of the colormap in Fig.~\ref{fig:channel-maps} were chosen to highlight the faint extended emission and to give a visual reference of the noise amplitude.
On top of it, we show contours at different levels of statistical significance.
All eight channel maps exhibit large patches of  $\geq 3\sigma$ emission distributed between \ang{;;0.2} and \ang{;;2.5} northward from the DSFG kinematic center (purple cross). 
In the first row of panels we can see that the plume grows longer at higher velocities, reaching a maximum isophotal extent of $\approx \ang{;;2.1} = \SI{13.8}{\kilo\parsec}$ at \SI{360}{\kilo\meter\per\second} (bottom left panel). 
In the subsequent channels, the plume becomes fainter and clumpy in appearance, possibly due to the lower S/N. 
We note that the DSFG centroid shifts coherently to the southeast as a result of rotation. 

\begin{figure*}
    \centering
    \includegraphics[width=17cm]{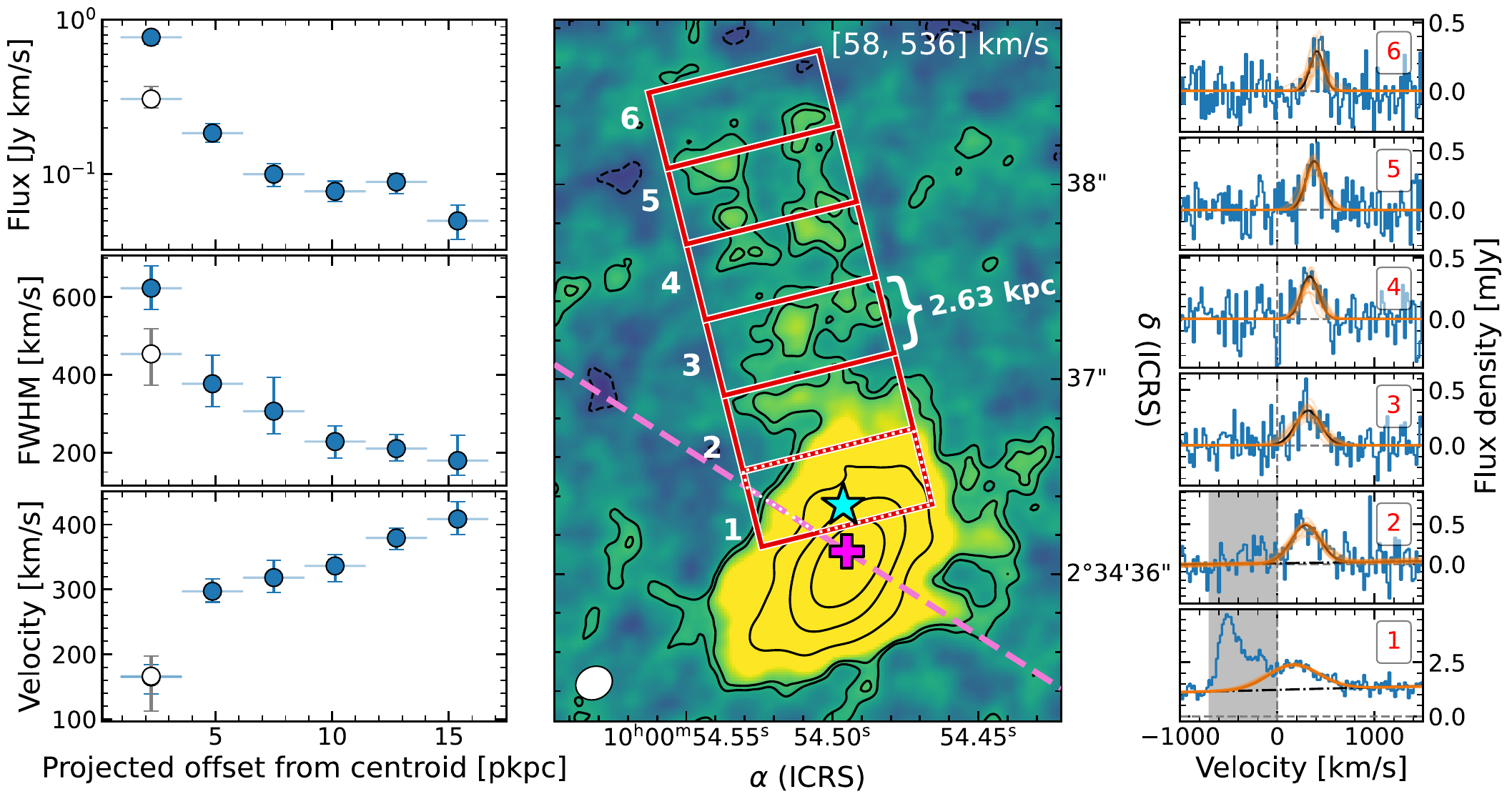}
    \caption{
    Spatial variation of spectral properties of the \cii plume.
    \textit{Left:} Radial profiles of fitted 1D Gaussian parameters in six apertures along the plume, as a function of projected distance from the kinematic center of the DSFG.
    The parameters shown are the integrated flux (top), the line FWHM (middle) and velocity centroid (bottom).
    White-filled markers in the first bin indicate best-fit values for the innermost spectrum after reducing contamination  with our symmetric difference method.
    \textit{Center:} \cii intensity map integrated from \SI{58}{\kilo\meter\per\second} to \SI{536}{\kilo\meter\per\second}, with black contours at the $\left\lbrace \pm2, 3, 15, 50, 150 \right\rbrace \times \sigma$  SB level. 
    Red rectangles numbered from 1 to 6 delineate the extraction apertures along the plume.
    The cyan star and pink cross mark the position of the rest-UV and the \cii kinematic centroids, respectively. The dashed line extrapolates the minor axis of \cb's \cii integrated emission at a PA=\ang{57.4} (see Sec. \ref{sec:results:morpho}). The line intersects the first rectangular aperture, defining a polygonal subregion where we extract the spectrum for symmetric difference analysis (see Appendix~\ref{sec:appendixA}).
    \textit{Right:} Extracted spectra from the six rectangular apertures, with their number labeled in red.
    The black solid line in each panel shows the maximum likelihood fit, while the orange lines are random samples from the posterior probability distribution.
    In panels 1 and 2 the shaded area indicates the velocity range excluded from the fit. The dash-dotted line shows the best-fit continuum component. 
    }\label{fig:radial_tail}
  \end{figure*}

We now want to quantify the spatial variations of the spectral profile along the plume with a proper treatment of the noise.
To this end, we place six adjacent rectangular $\ang{;;0.4}\times\ang{;;0.9}$ apertures covering the full extent of the plume as seen in Fig.~\ref{fig:radial_tail}, oriented with a position angle of \ang{14} east of north.
We extract the spectra from the Briggs-weighted (robust=0.5) datacube without continuum subtraction. With this weighting, the synthesized beam ($\ang{;;0.25}\times\ang{;;0.23}$) fits comfortably within each aperture, making them independent.

We then fit each extracted spectrum independently with a single 1D Gaussian, except for the first two apertures, where we include a first-order polynomial to model the continuum from the DSFG.
In these apertures we also mask emission from \SI{-700}{\kilo\meter\per\second} to \SI{0}{\kilo\meter\per\second} to avoid contamination from the approaching side of the DSFG.

Once again, we use \textsc{PyAutoFit} with the \textsc{Dynesty} \citep{Speagle2020Dynesty, Koposov2022DynestyZenodoV203} static nested sampler as a backend.
We assume a Gaussian likelihood for the sum of the rms-weighted residuals (data minus model).
For every channel (of width \SI{20}{\kilo\meter\per\second}), we measure the rms as the $3\sigma-$clipped standard deviation of the flux densities of 300 random apertures with the same size and orientation as the extraction apertures.
We adopt uniform priors for the three fitted parameters, namely the velocity center between \SI{-200}{\kilo\meter\per\second} and \SI{1000}{\kilo\meter\per\second}; the FWHM between \num{60} and \SI{1000}{\kilo\meter\per\second}; and the total flux between 0 and \SI{1.0}{\jansky\kilo\meter\per\second}.

Figure \ref{fig:radial_tail} and Table~\ref{tab:radial_values} show the results of these fits as a function of the distance of each aperture from the center of the DSFG.
We recognize radial trends in all the three parameters, and a large gap between apertures \#1 and \#2.
This apparent discontinuity could be due to additional contamination in aperture \#1's spectrum at positive velocities.
For this reason, we subtract a velocity-inverted spectrum from an aperture that mirrors aperture \#1 by the projected rotation axis (dashed line Fig.~\ref{fig:radial_tail}) and repeat the fit. After this correction, we find both lower fluxes and FWHM, but a consistent central velocity, as shown by the white-filled markers in Fig.~\ref{fig:radial_tail}. These differences illustrate the systematic uncertainties associated with aperture \#1, that make flux and FWHM less reliable.

Regardless of which fit we consider for the first aperture, we identify a radially decreasing trend for the flux density (or surface brightness). From \SI{5}{\kilo\parsec} to \SI{15}{\kilo\parsec} the flux drops almost exponantially, with an excess at $\approx$\SI{13}{\kilo\parsec}. Summing apertures from \#2 to \#6 yields a total flux of \SI{0.50+-0.04}{\jansky\kilo\meter\per\second}, which closely matches \ca's total flux (Table~\ref{tab:c1a_param}) and amounts to a \cii luminosity of \SI{3.1+-0.25e8}{\lumsol}. Adding aperture \#1 will raise this number by a factor of 2.53 under the original fitting scheme, and by a factor of 1.61 with the corrected fit.

In the middle panel of the left row we can see a decrease in the line width as one moves further out in the plume. Starting from $400-600\,\si{\kilo\meter\per\second}$ at \SI{2.5}{\kilo\parsec}, the plume narrows down to $180_{-38}^{+64}\,\si{\kilo\meter\per\second}$ in the outermost aperture.
We note that at low S/N the fits tend to bias the FWHM to larger values, hence the intrinsic FWHMs of the outermost bins might be even lower than depicted here. 

Our aperture-based measurements recover the velocity gradient that we had seen in the velocity map in Fig.~\ref{fig:maps}c. Here, both methods to extract the spectrum of aperture \#1 yield consistent central velocities at $v_\mathrm{cen}\approx\SI{180}{\kilo\meter\per\second}$. The next aperture (\#2) is already at \SI{300}{\kilo\meter\per\second}, suggesting a steep velocity gradient between apertures \#1 and \#2. In the subsequent apertures, the increase in velocity is less abrupt, going from $\approx \SI{300}{\kilo\meter\per\second}$ to $\approx \SI{400}{\kilo\meter\per\second}$ in the outer \SI{10}{\kilo\parsec}.

In addition, we create custom intensity maps for each aperture, integrating over a velocity window of size 1.1 times the best-fit spectral FWHM for maximal S/N. On these maps we obtain 1D spatial profiles along the long side of the corresponding  aperture, but extended to $\pm\SI{3}{\arcsec}$. We then fit a Gaussian to these profiles to obtain the transversal (spatial) FWHM of the plume. After deconvolving the beam width, we obtain FWHMs between \ang{;;0.4} and \ang{;;0.8} with an average of \ang{;;0.6}, but no clear radial trend.

Finally, we put upper limits on the FIR luminosity surface density of each aperture based on the continuum depth.
To this end, we randomly placed \num{600} rectangular apertures of size $\ang{;;0.9}\times\ang{;;0.4}$  on the rest-frame \SI{160}{\micro\meter}, \verb|robust=0.5| continuum map (without JvM correction).
We obtain a distribution of flux densities with $\sigma-$clipped standard deviation of $\sigma=\SI{38.8}{\micro\jansky}$.
Taking $5\sigma$ as the detection limit, we convert the flux density into FIR luminosity\footnote{Defined as the integral of the rest-frame SED between \SI{42}{\micro\meter} and \SI{122}{\micro\meter}.} assuming an underlying $T_\mathrm{dust}=\SI{45}{\kelvin}$ modified blackbody dust SED, with a dust emissivity index $\beta_\mathrm{dust}=1.5$ (a choice that roughly describes the average SED of star-forming galaxies at $z>4$, e.g., \citealt{Bethermin2020DataProcessing}). Such an SED gives the ratio $\nu L_\nu(\SI{158}{\micro\meter})/L_\mathrm{FIR} = 0.185$.
After correcting for primary beam gain decrement, we quote the resulting limits in Table~\ref{tab:radial_values}, except for aperture \#1 where emission from the DSFG dominates.
We note, however, that these values depend on the assumed temperature.
For example, choosing $T_\mathrm{dust}=\SI{55}{\kelvin}$ and $T_\mathrm{dust}=\SI{65}{\kelvin}$ produce $1.7\times$ and $2.4\times$ higher $L_\mathrm{FIR}$ limits, respectively. In any case, these measurements allow us to estimate the \cii/FIR diagnostic, which informs about the heating mechanism of the gas. At $T_\mathrm{dust}=\SI{45}{\kelvin}$ we obtain lower limits on the \cii/FIR ratio between $\approx$0.2\% (aperture \#6) and $\approx$0.6\% (aperture \#2). 
Unfortunately, these lower limits are not high enough to rule out UV photoelectric heating, characteristic of photodisocciation regions (PDRs; with typical \cii/FIR ratios between 0.01\% and 2\%, e.g., \citealt{HerreraCamus2018Shining1}), in favor of other mechanisms such as shock heating ($\gtrsim 4\%$, e.g., \citealt{Appleton2013ShockCiiSQ, Peterson2018TaffyCiiBridge}). Only at the coldest possible temperatures, $T_\mathrm{dust}\lesssim\SI{20}{\kelvin}$ ($T_\mathrm{CMB}=\SI{15}{\kelvin}$), the highest limit becomes \cii/FIR~$>6.8\%$, and thus harder to explain with UV photoelectric heating alone.

\begin{table*}
\renewcommand{\arraystretch}{1.5}
\centering
\caption{Extracted quantities from the apertures in Fig.~\ref{fig:radial_tail}}
\label{tab:radial_values}
\begin{tabular}{ccccccccc}	
	\hline
	Aperture & Flux density & $v_\mathrm{cen}$ & FWHM & $\Sigma_\mathrm{[CII]}$ & ${\Sigma_\mathrm{FIR}}$\tablefootmark{b} & $v_\mathrm{out}$ & $M_\mathrm{out}$ & $\dot{M}_\mathrm{out}$ \\
\# & \si{\jansky\kilo\meter\per\second} &  \si{\kilo\meter\per\second} & \si{\kilo\meter\per\second} & \SI{1e6}{\lumsol\per\kilo\parsec\squared} & \SI{1e10}{\lumsol\per\kilo\parsec\squared} & \si{\kilo\meter\per\second} & \SI{1e7}{\msun} & \si{\msun\per\year} \\
	\hline
1\tablefootmark{a} & $0.27_{-0.03}^{+0.05}$  & $166_{-54}^{+31}$ & $454_{-81}^{+65}$ & $10.6_{-1.3}^{+2.1}$ & - & \num{551+-75} & \num{25+-3} & \num{83+-18} \\
2 & $0.18_{-0.02}^{+0.03}$  & $297_{-17}^{+19}$ & $377_{-59}^{+73}$ & $7.3_{-0.9}^{+1.0}$ & $<1.3$ & \num{617+-56} & \num{17+-2} & \num{64+-11}\\
3 & $0.10_{-0.02}^{+0.02}$  & $318_{-23}^{+27}$ & $306_{-58}^{+86}$ & $4.0_{-0.7}^{+0.7}$ & $<1.3$ & \num{578+-66} & \num{9+-2}  & \num{32+-7}\\
4 & $0.08_{-0.01}^{+0.01}$ & $336_{-25}^{+17}$ & $228_{-42}^{+40}$ & $3.1_{-0.4}^{+0.5}$ & $<1.3$ & \num{530+-41} & \num{7+-1}  & \num{23+-4}\\
5 & $0.09_{-0.01}^{+0.01}$ & $379_{-18}^{+15}$ & $210_{-32}^{+36}$ & $3.5_{-0.6}^{+0.5}$ & $<1.3$ & \num{558+-33} & \num{8+-1}  & \num{28+-4} \\
6 & $0.05_{-0.01}^{+0.01}$  & $409_{-24}^{+27}$ & $180_{-38}^{+64}$ & $2.0_{-0.5}^{+0.5}$ & $<1.4$ & \num{561+-50} & \num{5+-1}  & \num{16+-4}\\
	\hline
	\hline
\end{tabular}
\tablefoot{From left to right, the columns display the aperture number, the flux density, the central line velocity, the FWHM, and the surface brigthness of \cii line emission, followed by the inferred maximum outflow velocity, gas mass, and mass outflow rates as described in Sec.~\ref{sec:discussion:outflow}.
\tablefoottext{a}{Values  in this row were extracted from the spectrum corrected via the  symmetric difference scheme (white filled circles in Fig.~\ref{fig:radial_tail}, see Appendix~\ref{sec:appendixA}).} \tablefoottext{b}{Limits on the far-infrared luminosity were estimated from the $5\sigma$ depth of the rest-frame \SI{160}{\micro\meter} continuum map, and assuming a modified blackbody SED of $T_\mathrm{dust}=\SI{45}{\kelvin}$ and $\beta_\mathrm{dust}=1.5$, which yields $\nu L_\nu(\SI{158}{\micro\meter})/L_\mathrm{FIR}=0.185$.}}
\end{table*}

\section{Discussion}\label{sec:discussion}
The main question we want to answer in this paper is what caused the \cii plume in the \sysname system. In this section we discuss possible scenarios such as outflows, inflows, ram pressure stripping, and tidal interactions.

\subsection{Conical outflow}\label{sec:discussion:outflow}

A first hypothesis on the nature of the \cii plume is to interpret it as a collimated outflow launched by \cb, driven by stellar feedback, AGN activity, or both.
The first case is compelling because \cb hosts a vigorous starburst in its center, which translates into a high rate of supernovae explosions that could potentially accelerate gas out of the galaxy.
The theoretical and observational evidence suggests that starburst-driven outflows escape along the minor axis of disk galaxies, because it is the path of least resistance, leading to (bi-)conical structures that extend perpendicular to the disk, with M82 \citep{BlandAndTully1988BipolarWindM82} and NGC 1482 \citep{VeilleuxAndRupke2002GalacticWindNgc1482} being archetypal examples in the local Universe. In contrast, AGN-driven outflows do not show a preferential alignment with the host minor axis \citep[e.g.,][]{Schmitt2003ExtendedOiiiSeyferts, RuschelDutra2021AgnifsOutflows} and can have very narrow opening angles \citep[$\theta\lesssim\ang{20}$, e.g][]{Sakamoto2014MergerWithMolecularOutflows, Aalto2020ResolvedMolecularJet}. 

In the J1000+0234 system, the \cii plume  diverges from the DSFG's minor axis by $\sim$\ang{40} clockwise, thus favoring an AGN origin.
Furthermore, the plume width remains narrow (transversal FWHM of $\approx$\ang{;;0.6}, see Sec.~\ref{sec:results:cii_spectral}) along its full extent, showing little to no broadening towards the top (North). In aperture \#6 of Fig.~\ref{fig:radial_tail}, the transversal FWHM is \ang{;;0.8}. Taking that as the opening of the cone at a distance of \ang{;;2.4} of its vertex at the center of the DSFG, we derive a projected angle of $\theta_\mathrm{p} = 2\arctan\left(0.4/2.4\right)\approx\ang{19}$, which is a projection of the true angle $\theta=2\times\arctan[\sin(i)\times\tan(\theta_\mathrm{p} / 2)]$.
Using the expectation value for a random uniform distribution of inclination angles, $\left<\sin{i}\right>=0.79$ (see the derivation in \citealt{Law2009HighRedshiftKinematics}), we estimate $\theta\approx\ang{15}$. This angle falls short of typical outflow opening angle observed in low redshift starbursts \citep[$\gtrsim\ang{60}$; e.g.,][]{HjelmAndLindblad1996OutflowNgc1365, Veilleux2001BiconicalOutflowNgc2992,  SeaquistAndClark2001CoInOutflowM82, Westmoquette2011SuperwindNgc253, Rubin2014EvidenceCollimatedOutflows} and in simulations \citep[e.g.,][]{Cooper2008ThreeDimSimWind, Nelson2019Tng50SimOutflows, Schneider2020SbDrivenOutflowsSim}.
This again favors AGN as the driver of the outflow, although a high gas density surrounding a central starburst can also lead to a strong collimation effect, especially in the molecular phase \citep[e.g.,][]{PereiraSantaella2016ExtendedMolecularOutflowLirg}.

On a side note, the fact we see only one cone instead of a symmetric bi-cone could be explained by power source being located above the midplane of the disk. In that case, the outflow would need to be much stronger to blow out into the other side.

Since \cii is the only line available, the dominant gas phase producing the emission remains unknown. 
One possibility is that most of the \cii flux arises from a population of clumps of cold (molecular and/or atomic) gas entrained in the hot outflowing plasma, as illustrated in the upper left panel of Fig.~\ref{fig:scenarios}. Such arrangement of the cold medium is predicted by the theory \citep[e.g.,][]{Schneider2020SbDrivenOutflowsSim, Kim2020MultiphaseOutflowsTigress, FieldingAndBryan2022MultiphaseGalacticWinds} and validated by a large set of observations \citep{Shapley2003RestUvSpectraLbgs, Rubin2014EvidenceCollimatedOutflows, PereiraSantaella2016ExtendedMolecularOutflowLirg, PereiraSantaella2018SpatiallyResolvedOutflowsUlirgs}.
Yet only a few studies exist reporting \cii outflows at high-$z$.
On one hand, high-velocity wings have been detected in individual \citep{HerreraCamus2021OutflowHZ4} and stacked spectra of main sequence galaxies \citep{Gallerani2018AlmaOutflowsRedshift5, Ginolfi2020AlpineOutflowsCGM}, and QSO hosts \citep[e.g.,][]{Bischetti2019QsoCiiOutflows}, but they usually come without any constraint on the state of the emitting gas.
On the other hand, \citet{Spilker2020MolecularOutflowsSPT} showed that 7 out of a sample of 11 lensed DSFGs host unambiguous molecular outflow signatures in the form of blueshifted OH \SI{119}{\micro\meter} absorption, yet none of these galaxies display broad \cii emission. These results led \citet{Spilker2020MolecularOutflowsSPT} to conclude that \cii line is an unreliable tracer of molecular outflows.
Since \cb has similar intrinsic properties as the sources in the \citet{Spilker2020MolecularOutflowsSPT} sample, we argue that it is unlikely for \cii to be tracing molecular gas in the plume. Instead, the gas could be composed of mainly atomic hydrogen.

Alternatively, the emission is dominated by ionized ($T\gtrsim\SI{1e4}{\kelvin}$) gas and we are seeing the outflow cone walls from the side, similar to what optical nebular lines show in low-redshift edge-on outflows.
When sufficient spatial resolution allows it, such cases exhibit a hallmark limb-brightening effect near the edges of the cone \citep[e.g.,][]{Strickland2004HighResXrayHalphaHotGasDisks, Westmoquette2011SuperwindNgc253, Venturi2017MagnumIonizedOutflows, Rupke2019AHundredKpcWindCgm, Herenz2023OutflowConeMetalPoor}.
Here, the plume is resolved in the transverse direction, but the S/N is too low to draw any conclusion about the structure. 
Nevertheless, \citetalias{Jimenez-Andrade2023LyaNebulaeGalaxyPair} detected \civ at the position of \cb, with a velocity shift comparable to what we measure in the plume ($\sim$\SI{300}{\kilo\meter\per\second}).
This would support the idea of a physical association of the \cii plume with warmer gas and hence a with a more energetic origin.

Regardless of the launching mechanism, we can now quantify some outflow properties based on the \cii line emission. In particular, we can estimate the mass of the plume enclosed in each of the apertures of Fig.~\ref{fig:radial_tail} from the measured \cii fluxes. In the optically thin limit, and assuming negligible background emission, the gas mass in each aperture is $M_\mathrm{out}=\kappa_\mathrm{[CII]}\times L_\mathrm{[CII]}$, where the conversion factor $\kappa_\mathrm{[CII]}$ depends strongly on the temperature, density and carbon abundance of the gas. Following \citet{HerreraCamus2021OutflowHZ4}, we assume the collisions are dominated by atomic hydrogen and adopt $\kappa_\mathrm{[CII]}=\SI{1.5}{\msun\per\lumsol}$ which corresponds to maximal excitation ($T\gg\SI{90}{\kelvin}$, $n\gg n_\mathrm{crit}\sim\SI{1e3}{\per\cubic\centi\meter}$), and solar abundance patterns. Since lower densities, temperatures and abundances yield higher values of $\kappa_\mathrm{[CII]}$, the masses we derive will effectively represent lower limits. For example, $\kappa_\mathrm{[CII]}$ becomes $3\times$ and $27\times$ larger when the metallicity drops to half solar, and one-tenth solar, respectively \citep[see discussion in][]{HerreraCamus2021OutflowHZ4}. 

In this way, we measure gas masses from \SI{5e7}{\msun} in the outermost aperture (\#6) to \SI{2.5e8}{\msun} in the innermost aperture (\#1, corrected by symmetric difference method). 
The sum of all the apertures gives \SI{7.1+-0.4e8}{\msun}.
We now exploit the spectral information to derive mass outflow rates as $\dot{M}_\mathrm{out} = M_\mathrm{out}\times v_\mathrm{out} \times \tan{(i)} / R $, where $v_\mathrm{out}$ represents the projected maximum outflow velocity and $R=\SI{2.63}{\kilo\parsec}$ is the projected length of the aperture in the short side (parallel to the flow). For the maximum outflow velocity we use the prescription of \citet{Genzel2011SinsSurvey} for outflows detected as broad emission components with Gaussian profile width $\sigma_\mathrm{broad}$ and velocity offset of $\left|\Delta v\right|$ with respect to the narrow component: $v_\mathrm{out}=\left|\Delta v\right| + 2\sigma_\mathrm{broad}$. Since we only detect a single spectral component along the plume, we take $\left|\Delta v\right|=v_\mathrm{cen}$ as the velocity centroid relative to the DSFG's systemic velocity. Both $v_\mathrm{out}$ and $M_\mathrm{out}$ values are listed in Table~\ref{tab:radial_values}.

Assuming $\left<i\right>=\ang{57.3}$, we obtain mass outflow rates between \SI{15}{\msun\per\year} and \SI{85}{\msun\per\year} (see last column of Table~\ref{tab:radial_values}). At face value, such an outflow will take several hundreds of mega-years to deplete the DSFG's gas reservoir of $\approx\SI{1e11}{\msun}$ \citep{Fraternali2021FastRotators}, and hence it's unlikely to quench the observed SFR anytime soon. We stress, however, that \cii conditions that differ from maximal excitation and solar abundance would drastically increase the inferred mass outflow rates. 

Mass outflow rates in starburst-driven winds scale with the SFR, and typically share the same order of magnitude \citep{Veilleux2020CoolOutflowsReview}.
When taken as lower limits, our fiducial mass outflow rates are roughly consistent with \cb's SFR given its large uncertainties.

Are these values consistent with AGN-driven outflows? Recalling that the X-ray luminosity of \cb is $L_\mathrm{2-10keV} < \SI{1.3e43}{\erg\per\second}$ \citep{Smolcic2015PhysicalPropertiesSmgsCosmos}, we estimate a bolometric luminosity of $L_\mathrm{bol} < \SI{1.3e44}{\erg\per\second}$ assuming a bolometric correction of 10, appropriate for low-luminosity broad line AGN \citep{VasudevanAndFabian2007BolCorrAGN}. According to the AGN wind scaling relations presented by \citet{Fiore2017AgnWindScaling}, our adopted $L_\mathrm{bol}$ upper limit allows for molecular outflows with mass outflow rates of up to \SI{100}{\msun\per\year} and maximum velocities of $\approx\SI{400}{\kilo\meter\per\second}$. In contrast, ionized outflows yield mass outflow rates of $\lesssim \SI{0.1}{\msun\per\year}$.
In summary, our mass outflow rates are consistent with the scaling relations for molecular outflows in both starbursts and AGN.

 The main caveat of the outflow interpretation is that it struggles to explain the kinematic structure of the \cii plume. 
 Specifically, most spatially resolved observations of galactic-scale outflows find a high velocity dispersion (FWHM $\gtrsim \SI{600}{\kilo\meter\per\second}$) with a flat, if not increasing, radial profile \citep[e.g.][]{Venturi2018MagnumNgc1365, Bao2019XshapedBicone, McPherson2023DuvetOutflowsMrk1486}. 
 This is also true in idealized high-resolution simulations such as the one presented by \citet{Schneider2020SbDrivenOutflowsSim}, since outflows that start as laminar, low-turbulence flows can develop and maintain instabilities as they interact with the CGM, thus increasing the velocity dispersion.
 These examples contrast with our measurement of radially decreasing line widths (middle panel of left row in Fig.~\ref{fig:radial_tail}).
 Also, if the emission is optically thin, the dispersion must increase with radius because the volume of the cone slice probed by the beam gets bigger, and thus includes a larger range of projected kinematic components.
 In other words, the dispersion should increase due to beam smearing even if the turbulence remains low.

  In conclusion, the outflow scenario is a natural explanation for the observed \cii plume, as it fits some of the expected properties an outflow would have given the nature of the DSFG. Future observations will be essential to rule out or confirm this scenario. For example, upcoming JWST/NIRSpec IFU observations will tell if there is any broad \halpha emission--tracer of ionized gas and a better-established indicator of outflows--associated with the \cii plume. In addition, deep rest-frame UV spectroscopy is needed to probe absorption by low-ionization metal species against the UV-bright regions of the system. Blueshifted lines will then unambiguously confirm the presence of cold outflows.

  \subsection{Gas accretion}\label{sec:discussion:inflow}
  We now consider a different possibility: the \cii plume traces a filament of inflowing gas. 
  Cosmological simulations have long predicted that massive halos at high-$z$ can be fed by narrow streams of cold gas, provided that the cooling timescale is shorter than the free-fall time \citep[e.g.,][]{DekelAndBirnboim2006ColdFlowsShockHeating}. 
  Moreover, if the galaxy at the center is a rotating disk,  it is expected that such streams reach the ISM along its minor axis. If the accretion is co-rotating with the disk, the disk structure is reinforced, whereas in the opposite case the disk might be disrupted \citep[e.g., ][]{Danovich2015AngularMomentumBuildUp}.

  As pointed out in Sec.~\ref{sec:results:masking}, the plume meets the DSFG at the receding end of the rotation, although it appears to overlap with the opposite side in the intensity map.
  This configuration resembles what could be observed in an accreting filament along the edge of a disk. 
  Such interpretation  would explain the relatively low-velocity dispersion, since cold streams are protected from virial shocks.

  In cosmological hydrodynamical simulations, cold gas flows in with a speed comparable to the virial velocity, and the speed is approximately uniform along the filaments \citep{GoerdtAndCeverino2015InflowVelocitiesColdStreams}.
  Following the estimation of \citet{Fraternali2021FastRotators} for the \sysname system, we assume a virial mass of $M_\mathrm{vir}=\SI{2e12}{\msun}$ and virial radius of $R_\mathrm{vir}=\SI{70}{\kilo\parsec}$, yielding a virial velocity of $V_\mathrm{vir}=\sqrt{M_\mathrm{vir}G/R_\mathrm{vir}}\approx\SI{350}{\kilo\meter\per\second}$. 
  \citet{GoerdtAndCeverino2015InflowVelocitiesColdStreams} found that halos of similar mass at $z \sim 4$ accrete at $V_\mathrm{stream}\approx 0.9 V_\mathrm{vir}$, so for \cb we would expect $V_\mathrm{stream}=\SI{315}{\kilo\meter\per\second}$ which is excellent agreement with our measured \cii velocities at 5-\SI{10}{\kilo\parsec}.
  Now, since these velocities are projected into the line of sight, the actual transversal speed could be much higher, depending on the inclination of the stream.
  Using again the average inclination for a random distribution of viewing angles, $\left<\sin{i}\right>=0.79$, we obtain $v/\left<\sin{i}\right> \approx \SI{400}{\kilo\meter\per\second}$. 
  This implies that the gas is moving faster than expected for a cold inflow stream. 
  We caution, however, that \citeauthor{GoerdtAndCeverino2015InflowVelocitiesColdStreams}'s simulations only track gas down to $0.2 R_\mathrm{vir}$, while the gas plume discussed here appears projected into smaller radii, from roughly zero to $0.2 R_\mathrm{vir}$, thus preventing a direct comparison. 

  If the gas retains or gradually loses angular momentum, it will follow a curved path as it falls. In that case, the observed velocity gradient can be due to a projection of a filament with relatively uniform gas inflow speed \citep[see][for a similar argument applied to the Abell 2390 central plume]{Rose2023Abell2390Plume}.
  But we observe the plume being approximately straight, so either the curvature is parallel to the line-of-sight or the gas is truly following a straight path. The latter seems more likely since it does not require a special orientation. So, if we assume the gas is falling straight into the galaxy, the positive velocity gradient we observe implies the gas is slowing down. This is contrary to the expectation of a free fall, where the gas will accelerate towards the center of the potential, but it is a plausible hydrodynamic effect where the pressure in the immediate vicinity of \cb exerts a force against falling gas.

  Since we detect the plume in \cii, the gas cannot be pristine but must have a significant mixture of processed material, even if the exact amount is not possible to constrain with the data in hand. Enrichment can occur by mixing of the infalling gas with the metals already present in the CGM thanks to past outflow activity. Alternatively, the gas is being enriched by star-formation ocurring in-situ. In fact, cosmological simulations also predict that accretion streams carry dwarf galaxies and star-forming clumps into the central halos \citep[e.g.,][]{Dekel2009ColdModeAccretion, Ceverino2010HighzClumpyDisks, Fumagalli2011AbsorptionInSimulatedColdStreams, Mandelker2018ColdFilamentaryAccretion}.   This idea has been recently proposed to explain the \SI{100}{\kilo\parsec}  filament of gas that feeds the massive radio-galaxy 4C 41.17, detected via its \ci emission \citep{Emonts2023CarbonCosmicStream}.
  To test this scenario we stacked the HST images from the four WFC3-IR filters available to search for associated UV sources (see Appendix~\ref{sec:appendixB}). We find two faint compact sources: one of them lies slightly outside the plume's \cii footprint to the North, while the other falls within the footprint at the the Eastern edge of aperture \#4, but only at S/N$\approx$3 (see Fig.~\ref{fig:hst_stack}). For this reason, we deem it a tentative detection and refrain from claiming a physical association with the plume. Otherwise, we do not detect UV emission  in the HST images at a stacked $5\sigma$ depth of \SI{26.2}{\mag\per\arcsec\squared}. This corresponds to an unobscured SFR density limit of roughly \SI{2.7}{\msun\per\year\per\kilo\parsec\squared}, assuming the \citet{Kennicutt98SFR} prescription scaled to the \citet{Chabrier2003IMF} IMF.

  Finally, a third possibility is that of gas recycling in a ``galactic fountain'', this is, gas ejected by the DSFG falls back in. However, recycled gas in cosmological simulations typically rains down in the outer parts of the disk, rather than in a collimated stream onto the center \citep[e.g.,][]{AnglesAlcazar2017BaryonCycleInFIRE, Grand2019AccretionAndFountainFlowsInAuriga}.

  In summary, interpreting the \cii plume as a cold accretion filament is qualitatively sound, although it requires some mechanism to slow down the gas as it falls, and a significant degree of metal enrichment. Theoretical predictions for the behavior of cold streams in the inner part of halos (below $0.2R_\mathrm{vir}$) will enable stronger conclusions.

  \subsection{Ram pressure stripping}\label{sec:discussion:rps}
  In certain conditions, galaxies lose their ISM through hydrodynamic interactions with their environment.
  Whenever a galaxy encounters a dense medium at high relative velocity, it will experience a drag force arising from the ram pressure acting on its ISM gas.
  If the drag force exceeds that of gravity, the gas becomes unbound.
  Known as ram pressure stripping \citep[RPS;][]{GunAndGott1972InfallClusters} this process is one of the main ways disk galaxies quench their star formation in galaxy clusters, but it can also be an efficient quenching mechanism for dwarf satellites around groups or individual massive galaxies \citep{Boselli2022RpsReview}.
  RPS in action is responsible for the existence of ``jellyfish'' galaxies, identifiable by their one-sided tails of stripped gas \citep[e.g.,][]{Chung2009VirgoVlaAtomicGas, Bekki2009RpsInDiscs, Smith2010UvTailsClusters, Ebeling2014Jellyfish, Poggianti2017GaspI}.

\begin{figure*}[!htb]
	  \centering
	  \includegraphics[width=15cm]{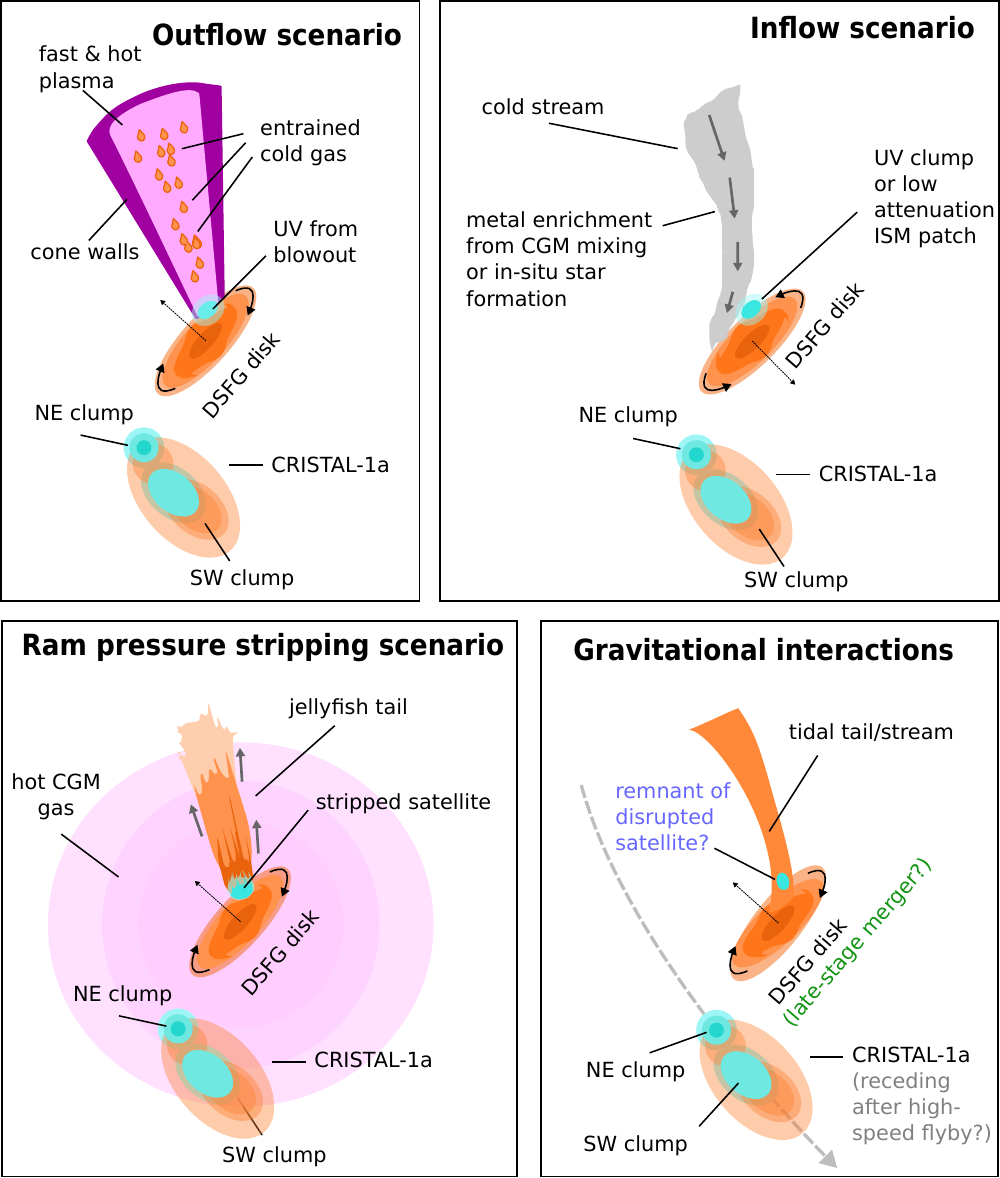}
	  \caption{Cartoon representations of the four possible scenarios to explain the \cii plume. In all panels the DSFG is depicted as an inclined rotating disk, with arrows showing the rotation.
	  The spiral arms and the normal vector are only intended to show the orientation of the disk, although we cannot determine it from the data.
  \textit{Upper left:} In the outflow scenario the extended \cii emission arises from clumps of cold gas entrained within a large-scale, off-axis conical wind.
Alternatively, \cii could be tracing ionized gas in the cone walls.
Here, the ISM at the launching site is blown out by the wind, allowing the escape of UV photons.
\textit{Upper right:} In the inflow scenario, a \cplus-bearing stream of gas falls from the north into the receding side of the disk. Here we depict the disk orientation inverted so that the inflow arrives in the near side of the disk.
\textit{Lower left:} In the ram pressure stripping scenario, the UV-bright clump in the DSFG is a satellite galaxy falling through a hot and dense halo around the system.
Ram pressure stripping of the satellite's ISM then produces the \cii plume in the form of a ``jellyfish'' tail.
\textit{Lower right:} In the gravitational interactions scenario, the plume is a tidal tail from a past gravitational interaction between the DSFG and (possibly) \ca. This cartoon summarizes the three cases discussed in the main text: a high-speed flyby of \ca (gray), a late-stage major merger (green), and the tidal disruption of a minor satellite (purple).
The dashed gray arrow qualitatively describes a possible orbit for \ca in the first case, that aligns with its morphological major axis.}\label{fig:scenarios}
  \end{figure*}

  In this section we discuss a scenario where the extended \cii feature is one of such tails (see lower left panel of Fig.~\ref{fig:scenarios}). Since the integrated \cii luminosity is comparable to that of the \ca galaxy, we could be seeing gas stripped from a galaxy with similar properties. In fact, the innermost tip of the feature closely matches the position of the UV-bright spot in the DSFG. While this light could be escaping through an opening in the dusty ISM of the DSFG (as suggested by its redder UV slope, $\beta_\mathrm{UV}\approx-1$; cf. Table~\ref{tab:glob_tab}), we cannot rule out it belongs to a separate smaller galaxy currently orbiting the DSFG. In this interpretation, the UV-bright region plus the \cii tail would represent the farthest example yet of a jellyfish galaxy.

How realistic is this scenario? RPS needs special conditions to be effective. Since the drag force is proportional to the density and the square of the relative velocity \citep[e.g.,][]{GunAndGott1972InfallClusters}, the medium has to be dense, volume-filling, and fast-moving.
Galaxy clusters are thus the ideal environment for RPS, since they are filled with a hot dense plasma (the intracluster medium, ICM, e.g., \citealt{Sarazin1986IntraclusterMedium}), and their huge gravitational potential allows members to acquire large orbital velocities.
The \sysname system is not a galaxy cluster, despite containing an excess number of LAEs and LBGs \citep{Smolcic2017SmgEnvironments, Jimenez-Andrade2023LyaNebulaeGalaxyPair}.
If anything, it could be classified as a protocluster, which means the structure is only loosely bound and cluster virialization has not happened yet \citep{Overzier2016ProtoclusterReview}.
Gas therefore has not reached the very high temperatures of typical clusters and is more likely dominated by small clouds and filaments of colder gas \citep[e.g.,][]{Hennawi2006ThickAbsorbersAroundQsos, Cantalupo2014CosmicWebFilament, ArrigoniBattaia2015CompactGasClumpsCgmQuasar}.
Nevertheless, \citetalias{Jimenez-Andrade2023LyaNebulaeGalaxyPair} reported extended \heii and \civ emission around the DSFG, with ratios (relative to \lya) that suggest strong ionization induced by AGN. The gas is then arguably warmer than $T=\SI{1e4}{\kelvin}$, increasing the cross section for a strong RPS interaction in a fast-passing satellite.

  Given the observed velocity difference $V$ between \cb and the base of the plume, we can estimate the minimum density the CGM must have in order to induce RPS. According to \citet{GunAndGott1972InfallClusters}, the requirement of ram pressure force being larger than the force holding the ISM of the satellite translates into the criterion
  \begin{equation}
	  \rho_\mathrm{CGM}V^2 > \frac{\Sigma_\mathrm{gas}v_\mathrm{rot}^2}{R_\mathrm{gal}},
  \end{equation}
  where $\Sigma_\mathrm{gas}$, $v_\mathrm{rot}$, and $R_\mathrm{gal}$ are the gas surface density, rotational velocity, and effective radius of the satellite, respectively, assuming a thin stellar disk structure. Using the inferred gas mass and aperture area of aperture \#1 (Table~\ref{tab:radial_values}), we adopt $\Sigma_\mathrm{gas}=\SI{1.6e7}{\msun\per\kilo\parsec\squared}$. Then, taking $v_\mathrm{rot}=\SI{100}{\kilo\meter\per\second}$, $R_\mathrm{gal}=\SI{200}{\parsec}$ and $V=\SI{166}{\kilo\meter\per\second}$, we obtain $n_\mathrm{CGM}=\rho_\mathrm{CGM}/m_{H}\gtrsim\SI{1}{\per\cubic\centi\meter}$. This is 3 to 4 orders of magnitude denser than typical ICM densities, and is comparable to the densities of \hii regions and the warm neutral medium in the Milky Way \citep{Draine2011IsmBook}. On one hand, these gas phases are normally clumpy in the Local Universe, so they would not fill enough volume to sustain effective RPS along \SI{15}{\kilo\parsec}. On the other hand, we know little about the structure of the high redshift CGM, so this concern may not apply.

The RPS scenario, however, is at odds with our observation of a decreasing linewidth as a function of distance along the plume.
In the context of galaxy clusters, evidence suggests that stripped cold gas interacts with the hot ICM either heating it or inducing instabilities that build up turbulence with time.
This effect is most clearly observed in ESO 137-001, a nearby edge-on jellyfish galaxy with a \SI{40}{\kilo\parsec} \halpha tail \citep{Sun2007Eso137HalphaTail}.
MUSE observations resolve the trailing diffuse \halpha emisison into three almost parallel tails, and all of them show a mild but significant increase in velocity dispersion in the direction away from the disk in the first \SI{20}{\kilo\parsec} \citep{Luo2023MuseKinematicsJellyfishEso137, Li2023TurbulenceJellyfish}. Beyond that distance the dispersion remains more or less uniform.
RPS in idealized numerical simulations generally reproduce this behavior, regardless of whether cooling \citep[e.g.,][]{RoedigerAndBruggen2008RamPressureStrippingWakes, Tonnesen2010RpsSignatures} and/or magnetic fields are included \citep[e.g.,][]{Tonnesen2014RpsMagneticFields}. 

  \subsection{Gravitational interactions}\label{sec:discussion:tidal}
  Finally, we consider the scenario where the plume is formed as the result of a gravitational disturbance.
  We know that \sysname is a complex multiple system (see Section~\ref{sec:backstory}), so we expect frequent interactions among its members.
  It is thus very plausible that a close encounter with the massive DSFG at the center of the group induced the formation of a tidal tail. 
  This explanation is simpler than the RPS scenario, as it does not require special hydrodynamic conditions but only gravity.

  Here we present a qualitative discussion of three different ways this interaction could have happened.
  First, we consider the case where \ca made a flyby close to \cb.
  Second, we assume the progenitors already merged and the plume is the lasting debris of such encounter.
  And third, we consider the plume as a separate dwarf galaxy altogether, currently in the process of being stripped by the tidal forces exerted by \cb.

  The idea that \ca and \cb are interacting was already suggested by several authors (e.g., \citealt{Capak2008ExtremeStarburst, Schinnerer2008MolGasMajorMerger}; \citetalias{GomezGuijarro2018AlmaMinorMergersSmgs}; \citetalias{Jimenez-Andrade2023LyaNebulaeGalaxyPair}), although it was mostly based on the short projected distance between the two objects. \citetalias{Jimenez-Andrade2023LyaNebulaeGalaxyPair} also took the UV and \lya elongated morphologies of \ca as evidence of ongoing tidal effects. 
  With reliable \cii-based redshifts at hand, we now know \ca moves at a projected speed of \SI{800}{\kilo\meter\per\second} relative to the DSFG, exceeding by a factor of \num{2.5} the escape velocity at a proper distance of \SI{10}{\kilo\parsec} (\SI{314}{\kilo\meter\per\second}; assuming a point mass of $M_\mathrm{dyn} = \SI{2.3e11}{\msun}$ located in the center of the DSFG, \citealt{Fraternali2021FastRotators}). This implies \ca is not gravitationally bound to \cb, indicating that if \ca actually interacted with \cb, it was in a high-speed flyby. 

  The main issue with this first tidal scenario is the mass ratio between \ca and \cb.
  Decades of work on numerical simulations of galaxy collisions have found that the largest and longest-lived tails are produced in ``major'' interactions \citep[i.e., where the progenitor mass ratio is 1:3 or lower; see][for a review]{DucAndRenaud2013TidesInCollidingGalaxiesReview}. In contrast, the stellar mass ratio here ($\sim 1:10$, \citetalias{GomezGuijarro2018AlmaMinorMergersSmgs}) belongs to the ``minor'' regime. Tidal features tend to be more prominent when the encounter occurs at low speeds, although it has been shown that high-speed flybys can also lead to the formation of gas-rich tails, provided the progenitors are both massive \citep[cf. the case of the VIRGOHI21 cloud in the Virgo cluster][]{Bekki2005GasTailVirgo21, DucAndBournaud2008Virgo21HighSpeedFlyby}. In conclusion, even if \ca shows some evidence of ongoing interaction with \cb, its relatively low mass make it an unlikely candidate to be responsible for the \cii plume.

  Based on these considerations, we now discuss the second scenario, where two massive progenitors already merged into \cb, ejecting the \cii plume as a tidal tail. The fact that \cb is rotation-supported does not necessarily rule out a merger origin, since simulations have shown that is possible for two massive galaxies to collide and result in a disk, provided they have high gas fractions \citep[e.g.,][]{SpringelAndHernquist2005FormationOfSpiralInMerger, Peschken2020DiscGalaxiesMergersIllustris}. In fact, this type of merger is expected to trigger very intense starbursts, just as the one we see in \cb.
 But did the merger have time to coalesce into a disk while still exhibiting a tidal tail? To answer this we need to estimate the dynamical age of the tail. To first order, we divide the length of the plume by the velocity difference between its two ends. In practice, we only consider apertures from \#2 to \#6, where the radial velocities are reliable. In this way we get $\tau_\mathrm{dyn}=\SI{13.15}{\kilo\parsec}/\SI{112}{\kilo\meter\per\second} = \SI{115}{\mega\year}$. On the other hand, the DSFG has a maximum rotation velocity of \SI{550}{\kilo\meter\per\second} at a radius of \SI{3.5}{\kilo\parsec} \citep{Fraternali2021FastRotators}, translating into a dynamical timescale of merely $\approx$\SI{40}{\mega\year}. Since the age of the plume is almost three times this value, we deem plausible that the tail persists after the merger has settled.

 Tidal tails in the Local Universe usually come in pairs, as in the well-known examples of the Antennae Galaxies or the Mice \citep[e.g.,][]{ToomreAndToomre1972TidalTailsAndBridges}. In contrast, \cb only shows one. While some examples of one-sided tails in late-stage mergers exist \citep[e.g., Mrk 273,][]{Sanders1988UlirgsAndOriginOfQsos}, their formation involves highly inclined encounters \citep{Howard1993TidalSimAtlas}. However, those configurations are not well suited for the survival of disks \citep[e.g.,][]{Cox2006KinMergerRemnants, Hopkins2013StarFormationMergers}.

 Finally, we consider the case of extreme tidal stripping of a satellite of \cb.
 In this picture, a dwarf galaxy orbiting \cb passes too close to the center, where tidal forces are strong enough to disrupt the intruder's ISM into a long stream of gas.
Depending on the central density of the intruder, it may not be totally disrupted, leaving its core relatively intact. If that is the case, the UV-bright region near \cb would be a natural candidate for the remnant, as it appears connected to the bottom of the plume. This idea could be tested with upcoming spectroscopy from JWST/NIRSpec.

Since gravity also acts on the stars, the three cases discussed here should produce a stellar stream associated with the plume, similar to those that populate the Milky Way halo \citep[e.g.][]{Malhan2018HaloStellarStreamsGaiaDr2} or the surroundings of nearby massive galaxies \citep[e.g.,][]{MartinezDelgado2023StellarStreamLegacySurvey}.
Here, as mentioned in Sec.~\ref{sec:discussion:inflow} we do not detect stellar emission from the plume in the HST images at a $5\sigma$ depth of \SI{26.2}{\mag\per\arcsec\squared}, corresponding to an intrinsic SB of \SI{18.8}{\mag\per\arcsec\squared} after correcting for cosmological dimming. 
 However, typical stellar streams in the Local Universe are much fainter, and only become detectable at sensitivities of $\approx\SI{28.5}{\mag\per\arcsec\squared}$ or higher \citep[e.g.,][]{BullockAndJohnston2005StellarHalosOne, MartinezDelgado2010StellarTidalStreamsSpirals}.
 In addition, our HST filters probe the rest-frame UV emission, and hence only trace young stars and not the bulk of the stellar mass. Upcoming JWST/NIRCam observations will be significantly deeper, and trace the rest-frame optical light. But even then, a detection would need the tail or stream to be remarkably bright.

 We conclude that gravitational interactions are more than capable of creating extended gas structures such as the \cii plume we observe, although the details are very uncertain. In fact, the three tidal scenarios presented here are only illustrative, and do not pretend to exhaust all the possible interactions that might produce a \cii plume. A more systematic approach would involve a series of numerical experiments to constrain the parameters that best reproduce the observed morphology and kinematics, but that goes beyond the scope of this paper.

\section{Summary \& conclusions}\label{sec:conclusions}
We have presented new ALMA Band 7 observations of the inner region of the \sysname system at $z=4.54$ located at the center of a bright \lya blob. These observations are part of the ALMA-CRISTAL Large Program, targeting \ciim emission and the underlying dust continuum in a sample of 25 star-forming galaxies at $4 < z < 6$.
The high sensitivity and angular resolution of these data reveal the detailed structure of the star-forming galaxies in the \sysname system. We report the discovery of a faint and diffuse \cii-emitting plume extending up to \ang{;;2.4} ($\approx$\SI{15}{\kilo\parsec}) from the central massive DSFG (\sysname-North). Complemented with archival MUSE and HST data, we analyzed the spatial and spectral properties of the plume and the two main galaxies of the system. Our main findings can be summarized as follows:

\begin{enumerate}
	\item The DSFG is detected in both \cii and dust continuum, and it shows a compact disk-like morphology. On one hand, dust emisison is fitted by a single 2D S\'ersic profile with circularized effective radius of $R_\mathrm{eff}\approx\SI{0.74}{\kilo\parsec}$, S\'ersic index $n\approx1.29$ and axis ratio $\approx0.4$. On the other hand, the \cii emission is fitted by 2D S\'ersic profile of $R_\mathrm{eff}\approx\SI{1.13}{\kilo\parsec}$, $n\approx0.7$ and axis ratio $\approx\num{.34}$.
	\item \ca sits at a projected distance of \ang{;;1.6} from the DSFG, and recedes \SI{800}{\kilo\meter\per\second} faster. It is detected in \cii emission but not in dust continuum, and its \cii morphology resembles that of the rest-frame UV emission. This means it is elongated and resolved into two clumps, although they are offset by \ang{;;0.3} from the corresponding UV-bright clumps. We model the northeastern clump with a circular exponential profile of $R_\mathrm{eff}\approx\SI{0.8}{\kilo\parsec}$, and the southwestern clump with an elliptical exponential of $R_\mathrm{eff}\approx\SI{2.4}{\kilo\parsec}$ and axis ratio $\approx0.82$.
	\item  The \cii plume starts at the center of the DSFG and extends northward with a position angle that is offset by \ang{40} clockwise from the DSFG's minor axis. The \cii surface brightness declines rapidly along the plume, becoming undetected at $\approx\SI{15}{\kilo\parsec}$ from the center of the DSFG.
  In the transverse direction we measure an average FWHM extent of $\approx\SI{4}{\kilo\parsec}$.
  \item The plume exhibits a clear velocity gradient, increasing the line central velocity as a function of radial distance from \SI{180}{\kilo\meter\per\second} to \SI{400}{\kilo\meter\per\second}, relative to the DSFG's systemic velocity. Moreover, the line FWHM also evolves with radius, showing a smooth drop from \SI{450}{\kilo\meter\per\second} at the center of the DSFG to \SI{190}{\kilo\meter\per\second} at the farthest measured point.
  \item We detect no dust continuum at rest-frame \SI{160}{\micro\meter} from the plume down to $5\sigma=\SI{194}{\micro\jansky}$ per $\ang{;;0.9}\times\ang{;;0.4}$ aperture. At an assumed dust temperature of $T_\mathrm{dust}=\SI{45}{\kelvin}$, we obtain lower limits on the \cii/FIR ratio between $\approx0.2\%$ and $\approx 0.6\%$, consistent with UV photoelectric heating of the gas.
  \item We estimate a minimum total mass for the plume of \SI{7.1+-0.4e8}{\msun}, asuming a conversion factor $\kappa_\mathrm{[CII]} = \SI{1.5}{\msun\per\lumsol}$ that corresponds to the limit of maximal excitation of the line in a medium dominated by atomic hydrogen, with solar-like carbon abundance.
\end{enumerate}

We discuss four scenarios to explain the results outlined above: (1), the plume is a conical outflow. (2), the plume traces a filament of inflowing gas. (3), the plume is a ram-pressure stripped tail of an infalling satellite. (4), the plume is tidal debris from past gravitational interactions. 

In the first scenario (1), we infer resolved mass outflow rates between \SI{16}{\msun\per\year} and \SI{83}{\msun\per\year}. The maximum outflow velocities across the plume range from \SI{530}{\kilo\meter\per\second} to \SI{620}{\kilo\meter\per\second}. These values are roughly consistent with literature scaling relations for SFR and $L_\mathrm{bol}$ in the case of starburst-driven and AGN-driven outflows, respectively. Given the very low ($\theta\approx\ang{15}$) opening angle we derive under the assumption of a simple conical geometry, plus the misalignment with the DSFG's minor axis, the putative outflow appears more likely to have originated in a central AGN. However, the observed kinematic radial trends are in mild tension with the expected properties of an outflow.

Scenario (2) is qualitatively consistent with theoretical expectations for cold accretion streams, except for our inference of a slow-down of the gas as it falls, although this can be accomodated by pressure gradients in the CGM. Moreover, the fact the putative filament emits in \cii rules out a chemically pristine gas composition. Still, simulations suggest that gas enrichment can happen concurrently during the inflow, either by gas mixing in the CGM or by in-situ star formation.

Scenario (3) requires the UV-bright region of the DSFG to be a satellite galaxy that is crossing the DSFG's CGM. In addition, the CGM must be dense ($\gtrsim \SI{1}{\per\cubic\centi\meter}$) in order to exert a significant ram pressure capable of stripping off the satellite's ISM. Moreover, observations and simulations of ram pressure stripping predict an increase of velocity dispersion along the stripped tails, contrary to what we observe in \sysname. 

Finally, scenario (4) is motivated by the fact \sysname is an overdense environment and close interactions must be frequent. We explored three possible ways a gravitational interaction can lead to the formation of a one-sided tidal tail. Namely, a high-speed flyby of \ca, a late-stage major merger in which the \cii plume is its remaining debris, and the tidal stripping of a minor satellite in a radial orbit. While a proper assessment of these configurations using tailored numerical simulations remains pending, heuristic arguments slightly disfavor the first two.

Besides scenario (3), which requires very special hydrodynamic conditions, outflows, inflows and tidal tails all have their pros and cons. Further observations and modeling are needed to discriminate between them. Promisingly, upcoming JWST/NIRCam and JWST/NIRSpec observations will deliver a high angular resolution view of the rest-frame optical morphology and spectral properties. For example, narrow \halpha spectral imaging with NIRSpec IFU will constrain the star formation rate density in the \cii plume. In addition, the detection of a broad \halpha component would support the outflow scenario. At the same time, NIRCam broad band observations will uncover the obscured regions of the DSFG and potentially discover a diffuse stellar stream associated with the \cii plume, providing further evidence to the tidal tail scenario.

Our results highlight the power of ALMA for characterizing the cold CGM in emission at high redshift, but also the difficulty of their interpretation. Disentangling the physical mechanisms that produce a given \cii observation is nevertheless very important for understanding how galaxies and their surroundings evolve.

\begin{acknowledgements}
  This paper makes use of the following ALMA data: ADS/JAO.ALMA\#2017.1.00428.L, ADS/JAO.ALMA\#2021.1.00280.L, ADS/JAO.ALMA\#2019.1.01587.S. ALMA is a partnership of ESO (representing its member states), NSF (USA) and NINS (Japan), together with NRC (Canada), MOST and ASIAA (Taiwan), and KASI (Republic of Korea), in cooperation with the Republic of Chile. The Joint ALMA Observatory is operated by ESO, AUI/NRAO and NAOJ.
  This paper is partly based on observations collected at the European Southern Observatory under ESO programs 0102.A-0448 and 0103.A-0272. M. S. was financially supported by Becas-ANID scolarship \#21221511, and also acknowledges ANID BASAL project FB210003. 
  M. A. acknowledges support from FONDECYT grant 1211951 and ANID BASAL project FB210003.
  A.F. acknowledges support from the ERC Advanced Grant INTERSTELLAR H2020/740120.
E. J. J. acknowledges support from FONDECYT Iniciaci\'on en investigaci\'on 2020 Project 11200263 and the ANID BASAL project FB210003.
I. DL. and S. vdG. acknowledge funding support from ERC starting grant 851622 DustOrigin.
K. T. acknowledges support from JSPS KAKENHI Grant Number 23K03466.
L. G. thanks support from FONDECYT regular proyecto No1230591.
R. I. is financially supported by Grants-in-Aid for Japan Society for the Promotion of Science (JSPS) Fellows (KAKENHI Grant Number 23KJ1006).
R. J. A. was supported by FONDECYT grant number 123171 and by the ANID BASAL project FB210003.
I.M. thanks the financial support by Grants-in-Aid for Japan Society for the Promotion of Science (JSPS) Fellows (KAKENHI Number 22KJ0821).
R. L. D. is supported by the Australian Research Council Centre of Excellence for All Sky Astrophysics in 3 Dimensions (ASTRO 3D), through project number CE170100013.
M. R. acknowledges support from project PID2020-114414GB-100, financed by MCIN/AEI/10.13039/501100011033.
\end{acknowledgements}

%
%

\bibliographystyle{aa}
\bibliography{ms.bib}

\begin{appendix}
  \section{Symmetric difference correction of aperture \#1's \cii spectrum}\label{sec:appendixA}	
  As described in Sec.~\ref{sec:results:cii_spectral}, the aperture \#1 in Fig.~\ref{fig:radial_tail} contains significant emission from the dust continuum and the approaching side of the DSFG rotator. To mitigate the contamination, we performed continuum subtraction and masked negative velocities before fitting the line, yet the resulting 1D Gaussian parameters deviate significantly from the parameters of the line in the subsequent apertures, suggesting additional contamination at positive velocities. Motivated by this, we attempt in this section to remove the additional emission before repeating the Gaussian fit.

  If the DSFG rotator is axially symmetric, we expect any aperture-extracted spectrum from one side of the DSFG to be similar--although with inverted velocities--to a spectrum extracted from a mirrored aperture, reflected with respect to the projected rotation axis. So in principle, differences in these spectra will uncover asymmetric emission in the cube, such as the plume we are studying. 

  Here, we take the projected rotation axis to be parallel to the minor axis of \cb's \cii emission (see dashed line in Fig.~\ref{fig:radial_tail}) and crossing the position of the steepest velocity gradient. In practice, we choose the point where the  2\textsuperscript{nd} moment (velocity dispersion) is maximal. Once we have defined the axis, we calculate the geometrical reflection of aperture \#1. However, the rotation axis crosses through the aperture, which would lead to an overlapping region between the aperture and its mirrored version. To avoid this, we crop the corner of aperture \#1 that goes over the eastern side of the axis before the reflection. 

\begin{figure}[!hbt]
    \centering
    \resizebox{\hsize}{!}{\includegraphics{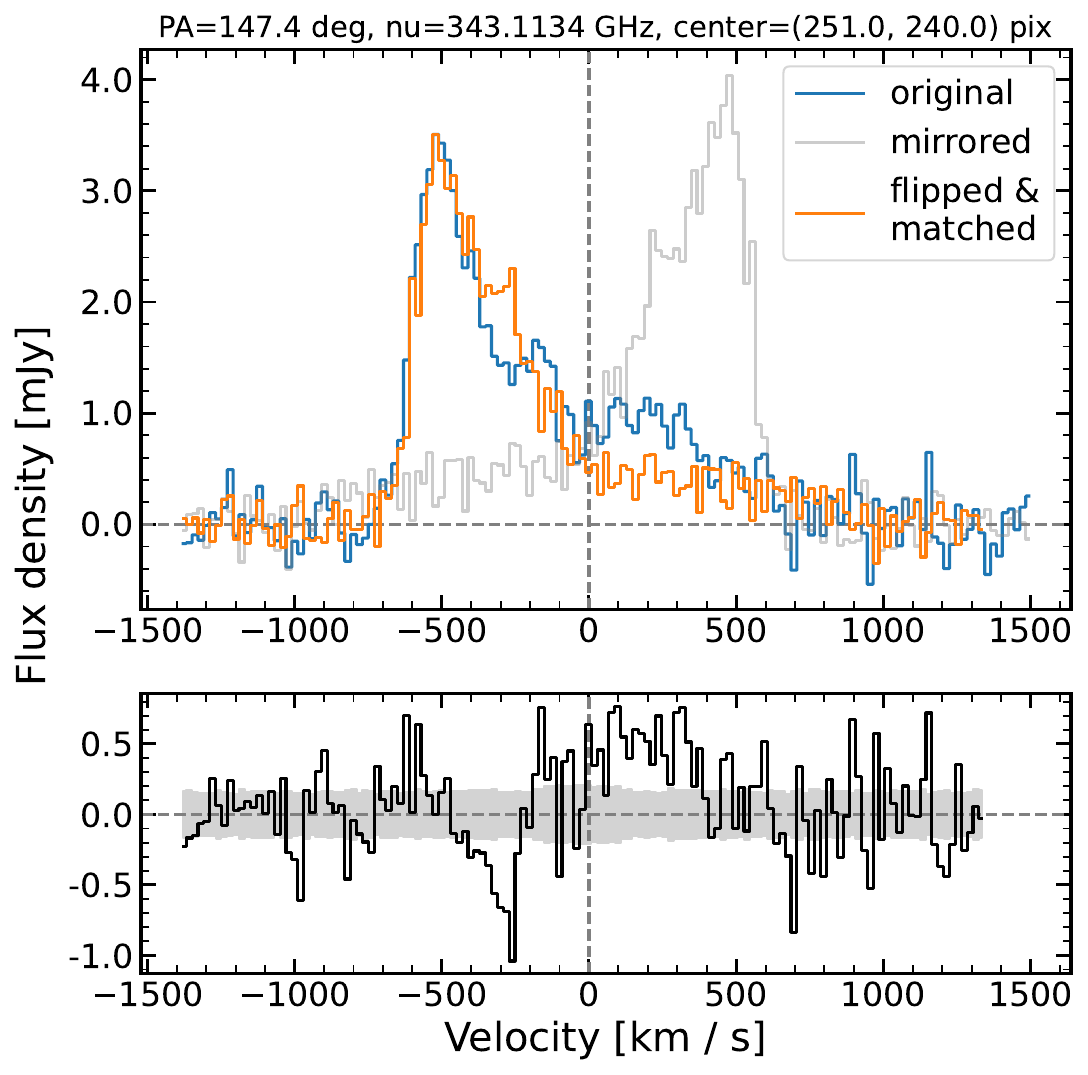}}	
    \caption{Symmetric difference analysis for aperture \#1. \textit{Upper panel:} \cii spectra of the cropped aperture \#1 (blue), its mirrored version (gray), and the mirrored version but velocity-inverted (flipped), shifted and matched to the original peak (orange). \textit{Lower panel:} Residuals from the difference between the original and the flipped-and-matched spectra.}\label{fig:sym_diff}
  \end{figure}

  We then extract the \cii spectrum from the \verb|robust=0.5|, \SI{20}{\kilo\meter\per\second} channel width cube without continuum subtraction, from both the cropped aperture \#1 and its reflection. We then apply a simple continuum subtraction by fitting a first-order polynomial to channels between $\pm\SI{1330}{\kilo\meter\per\second}$ and $\pm\SI{1000}{\kilo\meter\per\second}$ independently for each spectrum. We show the resulting spectra in Fig. \ref{fig:sym_diff}.

  After flipping the mirrored spectrum in velocity space and comparing it with the original, we find the horns have different peak heights and are offset by $\approx$\SI{47}{\kilo\meter\per\second}, but have similar overall shape. We thus shift and scale the mirrored spectrum to match the peak of the original and then perform the subtraction. The resulting spectrum is shown in the lower panel of Fig.~\ref{fig:sym_diff}. The difference is mostly consistent with zero, except for a strong but narrow negative difference around \SI{-250}{\kilo\meter\per\second} and a broad positive difference between \SI{0}{\kilo\meter\per\second} and \SI{400}{\kilo\meter\per\second}. The former arises from an excess or quirk in the spectrum from the mirrored aperture at \SI{+250}{\kilo\meter\per\second}, while the latter we assume to be associated with the \cii plume.

  Finally, we model the difference spectrum with the method described in Sec. \ref{sec:results:cii_spectral}. This is, we fit a 1D Gaussian using \textsc{PyAutoFit}. Once again, we mask the negative velocities down to \SI{-700}{\kilo\meter\per\second} to avoid fitting the residual negative feature. Compared to our fiducial method, this fit yields a significantly lower flux ($\approx\SI{0.3}{\jansky\kilo\meter\per\second}$) and FWHM ($\approx\SI{420}{\kilo\meter\per\second}$), although the central velocities are consistent. In conclusion, both the flux and FWHM in aperture \#1 are affected by contamination from the host, but the central velocity is not.

  \section{Stack of HST images}\label{sec:appendixB}
\begin{figure}
    \centering
    \resizebox{\hsize}{!}{\includegraphics{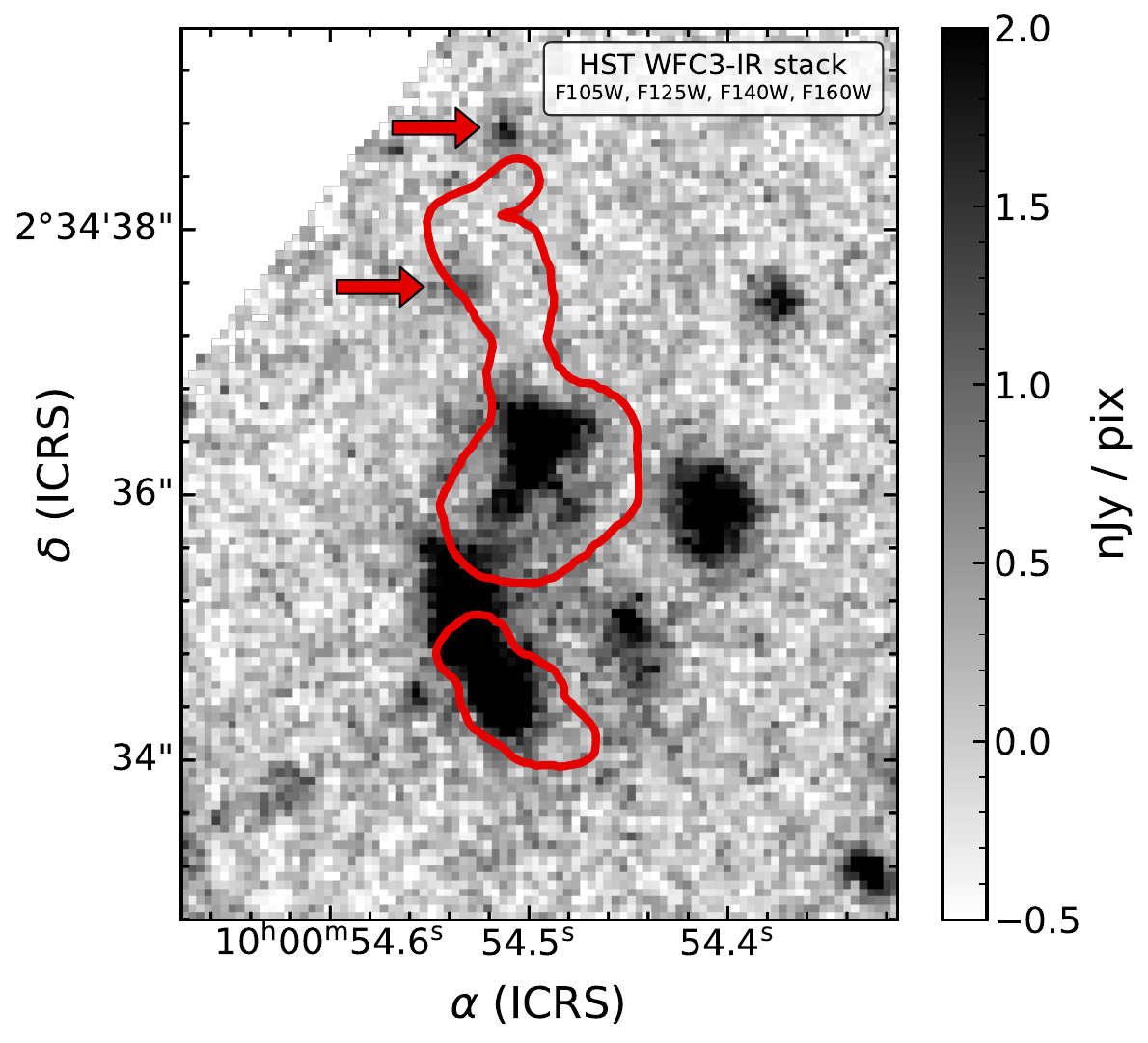}}	
    \caption{Stack of HST/WFC3-IR imaging of \sysname with the outline of the \cii emission in red. Arrows indicate the positions of the two sources potentially associated with the plume.}\label{fig:hst_stack}
  \end{figure}

  In this section we present the stack of HST imaging in the F105W, F125W, F140W and F160W filters. We converted the image units to nJy  per pixel, and then computed a pixel-by-pixel weighted sum. The weigths correspond to the inverse variance of the full frames. We then measure fluxes in 5000 random apertures of \SI{1}{\arcsec\squared} area, and obtain a $5\sigma$ depth of \SI{26.2}{\mag\per\arcsec\squared} Fig.~\ref{fig:hst_stack} shows the result of the stacking in the region close to the \sysname system, and reveals two tentative sources near the \cii plume that were not detected in the individual images.

\end{appendix}

\end{document}